\pdfoutput=1

\documentclass[a4paper,11pt]{article}
\usepackage{jheppub}

\usepackage{bm}

\usepackage{color}
\usepackage[usenames,dvipsnames,svgnames,table]{xcolor}

\usepackage{amsmath}
\usepackage{amssymb}
\usepackage{graphicx}
\usepackage{slashed}
\usepackage{soul}

\def\beq{\begin{equation}}
\def\eeq{\end{equation}}
\def\bea{\begin{eqnarray}}
\def\eea{\end{eqnarray}}
\def\ben{\begin{enumerate}}
\def\een{\end{enumerate}}

\def\lsim{\mathrel{\raise.3ex\hbox{$<$\kern-.75em\lower1ex\hbox{$\sim$}}}}
\def\gsim{\mathrel{\raise.3ex\hbox{$>$\kern-.75em\lower1ex\hbox{$\sim$}}}}
\def\ifmath#1{\relax\ifmmode #1\else $#1$\fi}

\title{Higgs Portals for Thermal Dark Matter
\\- EFT Perspectives and the NMSSM -}

\author[a,b]{Sebastian~Baum,}
\author[c,d]{Marcela~Carena,}
\author[e]{Nausheen~R.~Shah,}
\author[d,f]{Carlos~E.~M.~Wagner}

\affiliation[a]{The Oskar Klein Centre for Cosmoparticle Physics, Department of Physics, Stockholm University, Alba Nova, 10691 Stockholm, Sweden}
\affiliation[b]{Nordita, KTH Royal Institute of Technology and Stockholm University, Roslagstullsbacken 23, 10691 Stockholm, Sweden}
\affiliation[c]{Fermi National Accelerator Laboratory, P.~O.~Box 500, Batavia, IL 60510, USA}
\affiliation[d]{Enrico Fermi Institute and Kavli Institute for Cosmological Physics, University of Chicago, Chicago, IL 60637, USA}
\affiliation[e]{Department of Physics \& Astronomy, Wayne State University, Detroit, MI 48201, USA}
\affiliation[f]{HEP Division, Argonne National Laboratory, 9700 Cass Ave., Argonne, IL 60439, USA}

\emailAdd{sbaum@fysik.su.se}
\emailAdd{carena@fnal.gov}
\emailAdd{nausheen.shah@wayne.edu}
\emailAdd{cwagner@anl.gov}

\preprint{NORDITA-2017-130
\\\phantom{0} \hfill FERMILAB-PUB-17-611-T
\\\phantom{0} \hfill EFI-17-25
\\\phantom{0} \hfill WSU-HEP-1715}

\abstract{
We analyze a low energy effective model of Dark Matter in which the thermal relic density is provided by a singlet Majorana fermion which interacts with the Higgs fields via higher dimensional operators. Direct detection signatures may be reduced if blind spot solutions exist, which naturally appear in models with extended Higgs sectors. Explicit mass terms for the Majorana fermion can be forbidden by a $Z_3$ symmetry, which in addition leads to a reduction of the number of higher dimensional operators. Moreover, a weak scale mass for the Majorana fermion is naturally obtained from the vacuum expectation value of a scalar singlet field. The proper relic density may be obtained by the $s$-channel interchange of Higgs and gauge bosons, with the longitudinal mode of the $Z$ boson (the neutral Goldstone mode) playing a relevant role in the annihilation process. This model shares many properties with the Next-to-Minimal Supersymmetric extension of the Standard Model (NMSSM) with light singlinos and heavy scalar and gauge superpartners. In order to test the validity of the low energy effective field theory, we compare its predictions with those of the ultraviolet complete NMSSM. Extending our framework to include $Z_3$ neutral Majorana fermions, analogous to the bino in the NMSSM, we find the appearance of a new bino-singlino well tempered Dark Matter region.
}

\begin{document}

\maketitle
\flushbottom

\section{Introduction}
While the Standard Model~(SM) is extremely successful in describing the known particle interactions, it fails to explain the large scale structure of the Universe, since it does not provide a good Dark Matter~(DM) candidate. The simplest addition to the SM particle content would be in the form of a SM gauge singlet and in this work we shall concentrate on the particular example of a fermion as our DM candidate. In recent years, DM-nucleon scattering experiments such as LUX, XENON1T and PandaX-II have set stringent bounds on the possible couplings of DM to SM particles~\cite{Angloher:2015ewa,Agnese:2017jvy,Aprile:2017iyp,Cui:2017nnn,Akerib:2017kat,Amole:2017dex}. In particular, the coupling of the 125\,GeV Higgs boson to DM is significantly constrained. In addition, the vector coupling of DM to the Z gauge boson must be very small (see for instance Ref.~\cite{Escudero:2016gzx}), therefore we shall concentrate on singlet Majorana fermions, which couple only axially to the $Z$ boson. Such a fermion has the same gauge quantum numbers as a right-handed neutrino. One can define a matter parity, based on the $(B-L)$ quantum numbers of particles, namely $P = (-1)^{3 (B-L)}$, and demand interactions to be invariant under such a parity. Assuming that the DM carries no baryon $B$ or lepton $L$ numbers, this forbids all renormalizable interactions of the DM with SM particles, while allowing all SM Yukawa terms and Majorana masses for the right-handed neutrinos.

Since the coupling of DM to SM mediators is strongly constrained, we shall consider extending the SM by additional particles which can mediate interactions between DM and SM particles. Experimental precision tests of the SM strongly constrain extensions of the SM gauge sector, while far less is known about the SM Higgs sector. A well studied extension of the SM Higgs sector are so-called two Higgs doublet models (2HDMs)~\cite{Branco:2011iw}, which consist of adding a second Higgs doublet, as commonly found in models that provide a dynamical origin of the electroweak symmetry breaking~(EWSB) mechanism. The interactions of DM may quite generally be described by a set of non-renormalizable operators, including Majorana fermion bilinears and SM gauge invariant operators. The lower dimensional operators involve interactions with the Higgs fields and constitute a simple generalization of the so-called Higgs portal models~\cite{Kanemura:2010sh,Djouadi:2011aa,Lebedev:2011iq,Chang:2017gla}. As we shall see, the extended Higgs sector allows for the existence of \textit{blind spots} where the interaction of the Higgs bosons with DM particles may be reduced, satisfying direct detection constraints, while still allowing for the possibility of obtaining the observed (thermal) relic density~\cite{Perelstein:2012qg,Cheung:2012qy,
Huang:2014xua,Cheung:2014lqa,Han:2016qtc,Huang:2017kdh,Badziak:2017uto}. 

We shall require DM to be a weakly interacting massive particle (WIMP). In order to obtain a weak scale mass for the Majorana fermion in a natural way, we demand it to proceed from the vacuum expectation value~(vev) of a singlet scalar field, which develops in the process of EWSB. The absence of explicit masses may be the result of the presence of an explicit $Z_3$ symmetry, which also reduces the allowed number of higher dimensional operators and leads to a redefinition of the blind spot condition. 

A possible realization of these class of models is provided by supersymmetric~(SUSY) extensions of the SM~\cite{Martin:1997ns} which also allow for a dynamical explanation for the weak scale~\cite{Nilles:1983ge, Haber:1984rc, Martin:1997ns}. A particular virtue of the SUSY framework is that the stability of the Higgs mass parameter under quantum corrections can be ensured. In minimal extensions, the SM-like Higgs boson is naturally light~\cite{Casas:1994us, Haber:1996fp, Degrassi:2002fi}, and corrections to electroweak precision and flavor observables tend to be small, leading to good agreement with observations. Additionally, low scale SUSY leads to the unification of couplings at high energies and provides a natural DM candidate, namely the lightest neutralino. 

Among the simplest SUSY extensions, the Next-to-Minimal Supersymmetric extension of the SM~(NMSSM)~\cite{Ellwanger:2009dp}, fulfills all of the above properties while additionally containing a rich Higgs and neutralino spectrum. This may have an important impact on low energy observables. In particular, if the lighter neutralinos and neutral Higgs bosons are mainly singlets, they would be predominantly produced in association with heavier Higgs bosons or from the cascade decays of other SUSY particles, and therefore can easily avoid current direct experimental constraints~\cite{Kang:2013rj,King:2014xwa,Carena:2015moc,Ellwanger:2015uaz,Costa:2015llh,Baum:2017gbj,Ellwanger:2017skc}.

The light neutralino in the NMSSM is naturally mostly singlino-like, but with a non-negligible Higgsino component. Hence, its spin independent direct DM detection (SIDD) cross section is mediated predominantly via the SM-like Higgs boson. The current bounds on the SIDD cross section lead to relevant constraints on the couplings of DM to the SM-like Higgs boson, and demand the theory to be in the proximity of blind spots, where the contributions from the non-standard Higgs bosons become also relevant. The proper relic density may also be obtained; the thermal annihilation cross section is dominated by either resonant contributions of the Higgs bosons, or non-resonant $Z$ boson exchange contributions, with subdominant contributions from the light CP-even or the CP-odd Higgs bosons, the latter also having a large singlet component. In addition, a bino-like neutralino region with non-negligible Higgsino component may be present. In such a case, a sufficiently large thermal annihilation cross section yielding the proper relic density can be obtained by co-annihilation with the next-to-lightest neutralino, generally the singlino.
 
In Section~\ref{sec:EFT} we use the language of Effective Field Theories~(EFT) to outline the generic requirements of a model with singlet Majorana fermion DM and identify the required extended Higgs sector. In particular, we show the correlations of EFT parameters necessary to simultaneously obtain a thermal relic density, satisfy SIDD constraints, and accommodate a phenomenologically consistent Higgs sector. In Section~\ref{sec:NMSSM} we discuss the NMSSM as a possible ultraviolet completion of our EFT model, and demonstrate the mapping of EFT parameters to NMSSM parameters utilizing a top-down EFT approach. In Section~\ref{sec:DMpheno} we use the mature numerical tools available for the NMSSM to study the DM phenomenology, taking into account current collider and astrophysical constraints, as well as projections for the future. We identify two viable regions of parameter space with different DM phenomenology: 1) a new {\it well tempered} DM region, where the DM candidate is mostly bino-like and thermal production proceeds via resonant annihilation or co-annihilation with the singlino-like state, and 2) the region where the DM candidate is mostly singlino-like and the thermal relic density is mainly achieved via interactions mediated by the longitudinal mode of the $Z$ boson, the neutral Goldstone mode. Much of the phenomenology in both regions can be understood from the properties of the EFT worked out in Section~\ref{sec:EFT}, although some details are only found in complete models such as the NMSSM. We reserve Section~\ref{sec:Conclusions} for our conclusions. 

\section{An EFT for Singlet Dark Matter}\label{sec:EFT}

As motivated in the introduction, we will consider a model of SM singlet Majorana fermion DM, which has no renormalizable interactions with SM particles. In order to couple DM to the SM, we consider a 2HDM Higgs sector, more specifically, we shall take two Higgs doublets with opposite hypercharges $Y$, $H_u$ with $Y=+1/2$ and $H_d$ with $Y=-1/2$, which are naturally responsible for generating masses for the up and down-type quarks, respectively, as in type II 2HDMs. Since non-renormalizable interactions are suppressed by the scale associated with the masses of a heavy sector that was integrated out, one expects the dominant interactions to be associated with lower dimensional operators.

Including operators of dimension $d \leq 5$, the generic Lagrangian density describing interactions of a Majorana fermion $\chi$ with the two Higgs doublets $H_u$, $H_d$ is
\begin{equation} \label{eq:EFT1}
	{\cal L} = - \frac{\chi\chi}{\mu} \left[ \delta H_u \!\cdot\! H_d + \gamma (H_d^\dagger H_d + H_u^\dagger H_u) \right] - \frac{m_\chi}{2} \chi\chi 
	+ {\rm h.c.}~,
\end{equation}
where we have imposed a symmetry $H_d \leftrightarrow H_u$ and used a dot notation for $SU(2)$ products,
\begin{equation}
	H_u \!\cdot\! H_d = H_u^+ H_d^- - H_u^0 H_d^0~.
\end{equation}
Assuming, as usual, that both Higgs doublets acquire vevs, $\langle H_d \rangle = v_d$, $\langle H_u \rangle = v_u$, with $(v_d^2+v_u^2) = (174\mathrm{~GeV})^2$ and $\tan\beta = v_u/v_d$, we can define the Higgs basis~\cite{Georgi:1978ri, Donoghue:1978cj, gunion2008higgs, Lavoura:1994fv, Botella:1994cs, Branco99, Gunion:2002zf}\footnote{Note, that there are different conventions in the literature for the Higgs basis differing by an overall sign of $H^{\rm NSM}$ and $A^{\rm NSM}$.}
\begin{align}
	H^{\rm SM} &= \sqrt{2} {\rm Re} \left( \sin\beta H_u^0 + \cos\beta H_d^0 \right), \label{eq:Hbasis1}
	\\ G^0 &= \sqrt{2} {\rm Im} \left( \sin\beta H_u^0 - \cos\beta H_d^0 \right),
	\\ H^{\rm NSM} &= \sqrt{2} {\rm Re} \left( \cos\beta H_u^0 - \sin\beta H_d^0 \right),
	\\ A^{\rm NSM} &= \sqrt{2} {\rm Im} \left( \cos\beta H_u^0 + \sin\beta H_d^0 \right),
	\label{eq:Hbasis-1}
\end{align} 
where $H_d^0$ and $H_u^0$ denote the neutral components of the respective Higgs doublets. These may be related to the usual type II 2HDM Higgs bosons by the 
relations 
\begin{equation}
	H_d^i = \epsilon_{ij} H_1^{j *}\;, \;\;\;\;\;\;\;\;\;\;\;\;\; H_u^i = H_2^i\;.
\end{equation}

The $H^{\rm SM}$ interaction eigenstate has the same couplings to SM particles as a SM Higgs boson, $G^0$ is the (neutral) Goldstone mode making up the longitudinal polarization of the $Z$ boson after EWSB, and $H^{\rm NSM}$ and $A^{\rm NSM}$ are the non SM-like CP-even and CP-odd states, respectively. In particular, note that the Higgs basis fields are defined such that all the SM vev is acquired by the field corresponding to the neutral component of $H^{\rm SM}$, hence $\langle H^{\rm SM}\rangle = \sqrt{2} v$ and $\langle H^{\rm NSM}\rangle = 0$. Since the observed 125 GeV Higgs state $h$ appears to be close to SM-like in nature~\cite{Aad:2015zhl,Khachatryan:2016vau}, the interactions of $\chi$ with $h$ may be obtained from the above, approximating $h \sim H^{\rm SM}$ to first order. Ignoring the charged Higgs fluctuations, we obtain, at linear order in the fields,
\begin{equation}
	H_u \!\cdot\! H_d \to -\frac{v^2}{2} s_{2\beta} - \frac{v}{\sqrt{2}}\left( s_{2\beta} H^{\rm SM} + c_{2\beta} H^{\rm NSM} + iA^{\rm NSM}\right).
\end{equation}
Hence, the SM-like Higgs coupling to DM becomes
\begin{equation} \label{eq:gxxHSM}
	g_{\chi\chi h} \simeq g_{\chi\chi H^{\rm SM}} = \frac{\sqrt{2} v}{\mu} \left(\delta \sin 2\beta - 2 \gamma \right).
\end{equation} 
The interaction of a Majorana fermion with the SM-like Higgs boson listed above may be suppressed in three scenarios: 1) suppression of the couplings $\delta$ and $\gamma$; 2) large values of $\mu \gg v$; and 3) a particular correlation of the two couplings $\delta$ and $\gamma$ resulting in $g_{\chi\chi h}\sim0$. The last scenario, the so-called {\it blind spot} solution, is given by
\begin{equation} \label{eq:blindspot}
	\sin 2\beta = 2 \gamma/\delta\;.
\end{equation}

It is interesting to consider a model in which there are no explicit mass terms or scales and hence the Lagrangian is scale invariant. In such a situation, a natural way to generate the mass $m_\chi$ and the scale $\mu$ is via the vev of a singlet $S= \langle S \rangle+\frac{1}{\sqrt{2}} \left(H^S + i A^S \right)$. Hence, without loss of generality we can define $m_\chi = 2 \kappa\langle S \rangle$ and $\mu = \lambda \langle S \rangle$, where $\kappa$ and $\lambda$ are dimensionless parameters. 

The absence of explicit scale dependence could be understood as originating from a $Z_3$ symmetry, under which all scalar and fermion fields transform like $\Psi \rightarrow \exp[2 \pi i/3] \, \Psi$~(therefore also $\mu \rightarrow \exp[2 \pi i/3] \, \mu$). Besides forbidding explicit fermion mass terms, imposing such a $Z_3$ symmetry also forbids certain interactions. The remaining $d \leq 5$ terms are
\begin{equation}\label{eq:EFT}
	{\cal L} = - \frac{\chi\chi}{\mu} \left( \delta H_u \!\cdot\! H_d \right) - \kappa S \chi\chi + {\rm h.c.}\; ,
\end{equation}
resulting in the following DM-Higgs sector interactions:
\begin{equation} \begin{split} \label{eq:HCoup5}
	g_{\chi\chi H^{\rm SM}} = \frac{\sqrt{2} v}{\mu}\delta \sin 2\beta~, \qquad g_{\chi\chi H^{\rm NSM}} ~&= \frac{\sqrt{2} v}{\mu}\delta \cos 2\beta~, \qquad g_{\chi\chi A^{\rm NSM}} = i\frac{\sqrt{2} v}{\mu}\delta~, \\
 	g_{\chi\chi H^{\rm S}} ~&= i g_{\chi\chi A^{\rm S}} = -\sqrt{2} \kappa\;.
\end{split} \end{equation}

Imposing the $Z_3$ symmetry removes the possibility of a blind spot as defined in Eq.~(\ref{eq:blindspot}). The contributions from the $\chi\chi S$ coupling to either the thermal annihilation cross section relevant for the relic density or the SIDD cross section is further suppressed by singlet-doublet mixing since the singlet $S$ does not couple to SM particles beyond the Higgs sector. Hence, the dominant contributions to the SIDD and the thermal annihilation cross section will be proportional to $\delta^2$. Barring accidental cancellations between contributions from different Higgs bosons, the coupling $\delta$ must be suppressed in order to satisfy the stringent bounds from direct detection experiments. Hence, since current data implies that the dimension $d=5$ operators must be suppressed, we will include $d=6$ operators in the following. As we shall demonstrate, this will again allow for blind spot solutions to appear, enabling the suppression of the SIDD cross section. In addition, we find relevant contributions to the annihilation cross section from $d=6$ operators which will allow us to obtain sufficiently large annihilation cross sections to avoid over-closure of the Universe. The most relevant $d = 6$ operators are suppressed by powers of $m_{\chi}/\mu$ with respect to the $d=5$ ones, and thus become most relevant if the ratio $m_\chi/\mu$ is not very small. One could inquire about the impact of higher dimensional $d > 6$ operators in such a case. We shall address this question later by considering an ultraviolet completion of the EFT. Although the qualitative features found in the EFT remain valid in the complete theory, the precise quantitative predictions will indeed be affected to some degree by higher dimensional terms.
 
Assuming that the $d > 4$ terms in Eq.~(\ref{eq:EFT}) originate from a theory where a heavier $SU(2)$-doublet Dirac fermion with mass $\mu$ has been integrated out, we can write all the allowed $d=6$ operators which would arise from integrating out such a field. Ignoring the charged gauge boson interactions, we get 
\begin{equation} \begin{split} \label{eq:EFTmu}
	\mathcal{L} =&~ - \delta \frac{\chi\chi}{\mu} \left( H_u \!\cdot\! H_d \right)\left( 1 - \frac{\lambda \hat{S}}{\mu} \right) - \kappa S \chi\chi \left(1 + \xi \frac{H_d^\dagger H_d + H_u^\dagger H_u }{| \mu |^2} \right) + {\rm h.c.} \\
	&\qquad+ \frac{\alpha}{|\mu|^2}\left\{\chi^\dagger H_u^\dagger \bar{\sigma}^\mu \left[ i \partial_\mu - \frac{g_1}{s_W} (T_3 - Q s^2_W) Z_\mu \right] (\chi H_u) \right. \\
	& \qquad+ \left. \chi^\dagger H_d^\dagger \bar{\sigma}^\mu \left[ i \partial_\mu - \frac{g_1}{s_W} (T_3 - Q s^2_W) Z_\mu \right] (\chi H_d) \right\} ,
\end{split} \end{equation}
where $S = \mu/\lambda + \hat{S}$, $Q$ and $T_3$ are the charge and weak isospin operators, $s_W \equiv \sin \theta_W$ with the weak mixing angle $\theta_W$, and $g_1 = e/\cos \theta_W$ is the hypercharge coupling. Note, that the term proportional to $\hat{S}$ (the fluctuations of $S$) in the $\delta$-term arises because this originally $d=5$ term was actually suppressed by $1/\lambda S$, which we have expanded around the vev of $S$, yielding $1/\lambda S = 1/\mu - \lambda \hat{S}/\mu^2 + \mathcal{O}({S^2}/\mu^3)$. On the other hand, all the $d=6$ terms arising from integrating out a Dirac fermion are suppressed by $1/\lambda^2 S^{\dagger} S = 1/|\mu|^2 + \mathcal{O}(\hat{S}|/\mu^3)$ instead of $1/\lambda^2 S^2 = 1/\mu^2 + \mathcal{O}(\hat{S}|/\mu^3)$. Moreover, we have not included terms involving higher powers of the singlet field, since they are not expected to arise from integrating out a Dirac doublet fermion. The DM interactions with singlets are dominated by the tree level coupling $\kappa$, and the only modification from such terms would be a redefinition of the $S\chi\chi$ coupling $\kappa \to \kappa (1 + \mathcal{O}(m_\chi/\mu)$. Observe, that if dealing with on-shell $\chi$ fields, there is a redundancy in the above terms, since the application of the equation of motion on the terms proportional to the derivative of $\chi$ will lead to terms proportional to the $\chi$ mass, which also appear from the $\kappa$-term when inserting the vev of the field $S$. Another important point to note from Eq.~(\ref{eq:EFTmu}) is that the presence of derivative terms allows for interactions between the Goldstone $G^0$ and DM, absent in Eq.~(\ref{eq:EFT}), which as we shall see turn out to be relevant for the thermal annihilation cross section. For the convenience of the reader, we write Eq.~\eqref{eq:EFTmu} in terms of the Higgs basis states in the Appendix~\ref{app:L},~Eq.~\eqref{eq:EFTmuHbasis}.

From Eq.~(\ref{eq:EFTmu}), the coupling of the DM particles to the SM-like Higgs is given by
\begin{equation} \label{eq:gxxHSM1}
	g_{\chi\chi h} \simeq g_{\chi\chi H^{\rm SM}} = \frac{\sqrt{2} v}{\mu} \left[\delta \sin 2\beta - \frac{(\xi -\alpha)m_\chi}{ \mu^*} \right] ,
\end{equation} 
where the dependence on $\alpha m_\chi$ results from the application of the equations of motion. In general, we calculate the on-shell relationships by using the fact that, ignoring total derivatives,
\begin{equation} \label{eq:IbPandEoM}
	i (\partial_\mu \Phi ) \chi^\dagger i \bar{\sigma}^\mu \chi = - i \Phi \chi^\dagger i\bar{\sigma}^\mu ( \stackrel{\leftarrow}\partial_\mu + \stackrel{\rightarrow}{\partial_\mu}) \chi = i m_\chi \Phi \chi \chi + {\rm h.c.}~,
\end{equation}
where $\Phi$ is a real scalar field. Note, that the direct expansion of the derivative terms proportional to $\alpha$ in Eq.~\eqref{eq:EFTmu} leads to interactions with the CP-even Higgs bosons when the derivative is acting on the Majorana fermion fields, and to derivative interactions with the CP-odd Higgs states when the derivative is acting on the Higgs doublets, as required by hermiticity.

We see that the blind spot for the cancellation of the coupling of $H^{\rm SM}$ to pairs of DM now occurs for
\begin{equation}\label{eq:bs}
	\sin 2 \beta = \frac{(\xi-\alpha) m_\chi}{\mu^* \delta}\;,
\end{equation}
and we can further match the interactions dictated by the Lagrangians given in Eqs.~(\ref{eq:EFT1}) and (\ref{eq:EFTmu}) by noting that
\begin{equation}
	\gamma = \frac{(\xi -\alpha)m_\chi}{2 \mu^*}\;.
\end{equation}

The $\chi$ interactions with $H^{\rm NSM}$ are 
\begin{equation} \label{eq:gxxHNSM}
	g_{\chi\chi H^{\rm NSM}} = \frac{\sqrt{2} v}{\mu} \delta \cos 2\beta \;.
\end{equation}
Note, that there are no terms proportional to $\gamma$ (or $m_\chi$) and therefore there is no blindspot such as the one for $H^{\rm SM}$ in Eq.~(\ref{eq:bs}); instead $g_{\chi\chi H^{\rm NSM}}\to 0$ for $\tan\beta \to 1$.

On the other hand, the interactions with the CP-even singlet state are given by
\begin{equation} \label{eq:gxxHS}
	g_{\chi \chi H^S} = - \sqrt{2} \left\{\frac{v^2}{2\mu^2} \delta \lambda \sin2\beta + \kappa \left[ 1 + \frac{(\xi-\alpha) v^2}{|\mu|^2} \right]\right\}. 
\end{equation} 
Here, the dependence on $\alpha$ comes from a field renormalization of $\chi$ necessary to retain a canonical kinetic term for $\chi$ when including dimension $d\leq6$ operators. In principle, this field renormalization introduces corrections to all couplings of $\chi$. However, we are only considering operators of $d \leq 6$. The modification from the field renormalization is suppressed by $|\mu|^{-2}$, hence, this correction is only relevant for the renormalizable $\chi\chi\hat{S}$ interactions.

The interactions of $\chi$ with the CP-odd scalars are easy to read from the above as well. For instance, although the Goldstone interactions involve derivatives of the Goldstone fields, for on-shell $\chi$'s one can use Eq.~\eqref{eq:IbPandEoM} to obtain the interaction with the (neutral) Goldstone mode
\begin{equation} \label{eq:gxxG0}
	g_{\chi \chi G^0} = - i \frac{\sqrt{2} m_\chi v}{|\mu|^2} \alpha \cos2\beta\;.
\end{equation}
The orthogonal state, $A^{\rm NSM}$, also has relevant interactions with DM, namely 
\begin{equation} \label{eq:gxxA}
	g_{\chi\chi A^{\rm NSM}} = i \frac{\sqrt{2} v}{\mu} \left(\delta + \frac{m_\chi }{ \mu^*} \alpha \sin2\beta\right). 
\end{equation}
Finally, the interactions of the CP-odd singlet state $A^S$ are analogous to its CP-even counterpart,
\begin{equation} \label{eq:gxxAS}
	g_{\chi \chi A^S} = i \sqrt{2} \left\{\frac{v^2}{2\mu^2} \delta \lambda \sin2\beta + \kappa \left[ 1 + \frac{(\xi-\alpha) v^2}{|\mu|^2} \right]\right\}. 
\end{equation} 

\subsection{Higgs Sector} \label{sec:EFT_Higgs}

In the previous section we have motivated a structure for the scalar sector consisting of two Higgs doublets and one singlet, all three of which acquire a vev. We can define the {\it extended Higgs Basis}, $\{ H^{\rm SM}, H^{\rm NSM}, H^S\}$ for the CP-even states and $\{A^{\rm NSM}, A^S\}$ for the CP-odd states, where the doublet components are as defined in Eqs.~\eqref{eq:Hbasis1}--\eqref{eq:Hbasis-1}, and the singlet $S = \langle S\rangle + \hat{S} = \mu/\lambda + \frac{1}{\sqrt{2}} \left( H^S + i A^S \right)$ does not get rotated~\cite{Carena:2015moc}. These interaction eigenstates mix into mass eigenstates. We denote the CP-even mass eigenstates as $h_i = \{ h, H, h_S \}$,
\begin{equation} \label{eq:hi}
	h_i = S_{h_i}^{\rm SM} H^{\rm SM} + S_{h_i}^{\rm NSM} H^{\rm NSM} + S_{h_i}^S H^S \;,
\end{equation}
and the CP-odd states as $a_i = \{A, a_S\}$,
\begin{equation} \label{eq:ai}
	a_i = P_{a_i}^{\rm NSM} A^{\rm NSM} + P_{a_i}^S A^S \;.
\end{equation}
The mixing angles $S_i^j$ and $P_i^j$ are obtained from the diagonalization of the corresponding mass matrices. We can write the (symmetric) squared mass matrix of the CP-even Higgs bosons in the extended Higgs Basis as
\begin{equation}
	\mathcal{M}_S^2 = \begin{pmatrix} \mathcal{M}_{S,11}^2 & \mathcal{M}_{S,12}^2 & \mathcal{M}_{S,13}^2
	\\ \mathcal{M}_{S,12}^2 & \mathcal{M}_{S,22}^2 & \mathcal{M}_{S,23}^2
	\\ \mathcal{M}_{S,13}^2 & \mathcal{M}_{S,23}^2 & \mathcal{M}_{S,33}^2 \end{pmatrix}.
\end{equation}
Since the observed Higgs boson is predominantly SM-like, we can parametrize the elements corresponding to $H^{\rm SM}$--$H^{\rm NSM}$ and $H^{\rm SM}$--$H^S$ mixing as
\begin{equation}
	\mathcal{M}_{S,12}^2 \equiv \epsilon \overline{\textbf{M}}_{S,12}^2 \;, \qquad \mathcal{M}_{S,13}^2 \equiv \eta \overline{\textbf{M}}_{S,13}^2 \;,
\end{equation}
where $\epsilon$ and $\eta$ are small parameters $\epsilon, \eta \ll 1$, and we have defined $\overline{\textbf{M}}_{S,12}^2 \equiv \sqrt{\mathcal{M}_{S,11}^2 \mathcal{M}_{S,22}^2}$, and similarly, $\overline{\textbf{M}}_{S,13}^2 \equiv \sqrt{\mathcal{M}_{S,11}^2 \mathcal{M}_{S,33}^2}$. Barring the possibility of very degenerate diagonal mass terms, these relations ensure that the SM-like state has only small mixings with the non-standard states and that its mass squared can be approximately identified with the $\mathcal{M}_{S,11}^2$ matrix element. Observe, that after imposing the minimization conditions $\mathcal{M}_{S,11}^2$ and $\mathcal{M}_{S,12}^2 = \varepsilon \overline{\textbf{M}}_{S,12}^2$ become proportional to the square of the Higgs vev $v$. The matrix elements $\mathcal{M}_{S,13}^2 = \eta \overline{\textbf{M}}_{S,13}^2$ and $\mathcal{M}_{S,23}^2$ are only linear in $v$. 

Keeping terms to linear order in the small parameters $\varepsilon$ and $\eta$ only, the eigenvalues are
\begin{equation}
	m_h^2 \simeq \mathcal{M}_{S,11}^2~, \qquad m_{h_S,H}^2 \simeq \frac{\mathcal{M}_{S,22}^2 + \mathcal{M}_{S,33}^2 \mp \sqrt{\left(\mathcal{M}_{S,22}^2-\mathcal{M}_{S,33}^2\right)^2 + 4 \left(\mathcal{M}_{S,23}^2\right)^2} }{2}~.
\end{equation}
The eigenvectors are
\begin{align}
	\frac{S_{h}^{\rm NSM}}{S_h^{\rm SM}} &= \frac{-\eta \overline{\textbf{M}}_{S,13}^2 \mathcal{M}_{S,23}^2 - \epsilon \overline{\textbf{M}}_{S,12}^2 \left(m_h^2 - \mathcal{M}_{S,33}^2\right)}{\left(\mathcal{M}_{S,23}^2\right)^2 - \left(m_h^2 - \mathcal{M}_{S,22}^2\right) \left(m_h^2-\mathcal{M}_{S,33}^2\right)}\;, \\
	\frac{S_h^S}{S_h^{\rm SM}} &= \frac{ - \epsilon\overline{\textbf{M}}_{S,12}^2 \mathcal{M}_{S,23}^2 - \eta \overline{\textbf{M}}_{S,13}^2 \left( m_h^2 - \mathcal{M}_{S,22}^2 \right)}{\left(\mathcal{M}_{S,23}^2\right)^2 - \left(m_h^2 - \mathcal{M}_{S,22}^2\right) \left(m_h^2-\mathcal{M}_{S,33}^2\right)}\;,
\end{align}
for the SM-like mass eigenstate,
\begin{align}
	\frac{S_H^{\rm SM}}{S_H^{\rm NSM}} &= \frac{ -\eta \overline{\textbf{M}}_{S,13}^2 \mathcal{M}_{S,23}^2 - \epsilon \overline{\textbf{M}}_{S,12}^2 \left( m_H^2 - \mathcal{M}_{S,33}^2 \right) }{ \eta^2 \left( \overline{\textbf{M}}_{S,13}^2\right)^2 - \left( m_H^2 - \mathcal{M}_{S,11}^2 \right) \left( m_H^2 - \mathcal{M}_{S,33}^2\right) }\;, \\
	\frac{S_H^S}{S_H^{\rm NSM}} &= \frac{ - \epsilon \eta \overline{\textbf{M}}_{S,12}^2 \overline{\textbf{M}}_{S,13}^2 - \mathcal{M}_{S,23}^2 \left( m_H^2 - \mathcal{M}_{S,11}^2 \right) }{ \eta^2 \left( \overline{\textbf{M}}_{S,13}^2 \right)^2 - \left( m_H^2 - \mathcal{M}_{S,11}^2 \right) \left( m_H^2 - \mathcal{M}_{S,33}^2 \right) }\;,
\end{align}
for the doublet-like eigenstate, and
\begin{align}
	\frac{S_{h_S}^{\rm SM}}{S_{h_S}^S} &= \frac{ - \epsilon \overline{\textbf{M}}_{S,12}^2 \mathcal{M}_{S,23}^2 - \eta \overline{\textbf{M}}_{S,13}^2 \left( m_{h_S}^2 - \mathcal{M}_{S,22}^2 \right) }{ \epsilon^2 \left( \overline{\textbf{M}}_{S,12}^2 \right)^2 - \left( m_{h_S}^2 - \mathcal{M}_{S,11}^2 \right) \left( m_{h_S}^2 - \mathcal{M}_{S,22}^2 \right) } \;, \\
	\frac{S_{h_S}^{\rm NSM}}{S_{h_S}^S} &= \frac{ - \epsilon \eta \overline{\textbf{M}}_{S,12}^2 \overline{\textbf{M}}_{S,13}^2 - \mathcal{M}_{S,23}^2 \left( m_{h_S}^2 - \mathcal{M}_{S,11}^2 \right) }{ \epsilon^2 \left( \overline{\textbf{M}}_{S,12}^2 \right)^2 - \left( m_{h_S}^2 - \mathcal{M}_{S,11}^2 \right) \left( m_{h_S}^2 - \mathcal{M}_{S,22}^2 \right) }\,,
\end{align}
for the singlet-like mass eigenstate.

If we use the approximate eigenmasses, we find for the SM-like mass eigenstate
\begin{equation}
	S_h^{\rm SM} \approx 1 \;, \qquad \frac{S_h^{\rm NSM}}{S_h^{\rm SM}}, \frac{S_h^S}{S_h^{\rm SM}} = \mathcal{O}(\epsilon, \eta) \;,
\end{equation}
and for the other mass eigenstates
\begin{align}
	S_H^{\rm SM} &\approx S_{h_S}^{\rm SM} \approx 0 ~, \\ 
	-\frac{S_H^S}{S_H^{\rm NSM}} &\approx \frac{S_{h_S}^{\rm NSM}}{S_{h_S}^S} \approx \frac{2 \mathcal{M}_{S,23}^2}{\mathcal{M}_{S,22}^2 - \mathcal{M}_{S,33}^2 + \sqrt{ \left( \mathcal{M}_{S,22}^2 - \mathcal{M}_{S,33}^2 \right)^2 + 4 \left(\mathcal{M}_{S,23}^2\right)^2} }~, \\
	S_H^{\rm NSM} & \approx S_{h_S}^S \approx \left[1 +\left( \frac{S_{h_S}^{\rm NSM}}{S_{h_S}^S} \right)^2 \right]^{-1/2}~.
\end{align}

\subsection{EFT: Relic Density} \label{sec:EFT_relic}

\begin{figure}
	\begin{center}
		\includegraphics[height=3.8cm]{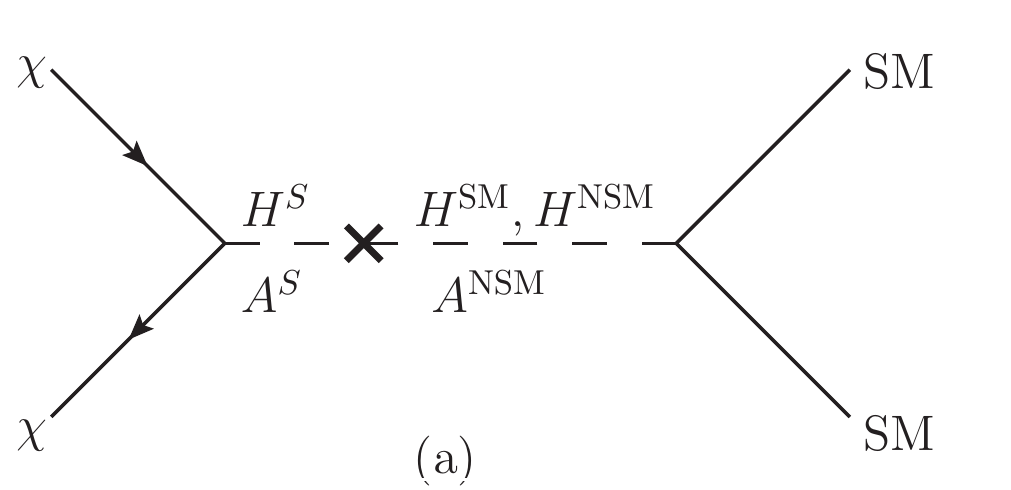}
		\hfill
		\includegraphics[height=3.8cm]{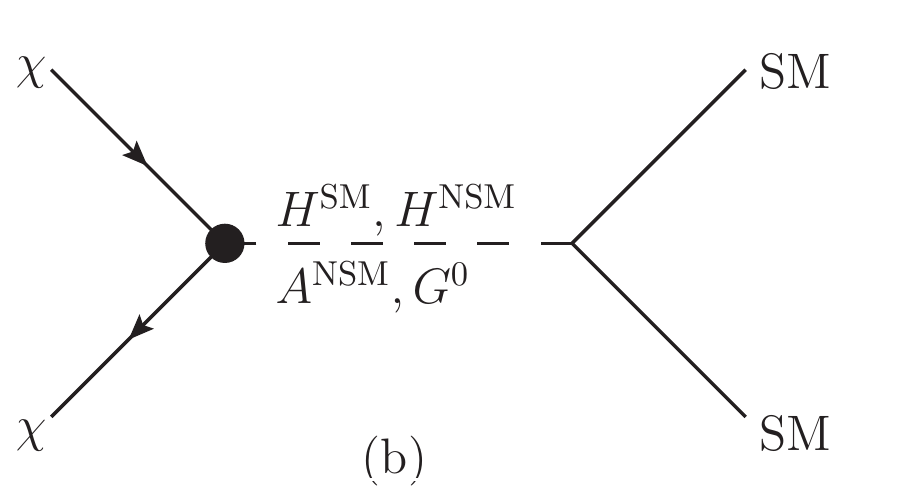}
		\caption{Exemplary diagrams illustrating the interactions of singlet DM $\chi$ with SM particles. The left diagram (a) depicts interactions arising via the tree-level interaction of a pair of DM singlets $\chi$ with the scalar singlet states $H^S$ and $A^S$, which mix (indicated by the cross) with the Higgs basis states from the Higgs doublets $H^{\rm SM}, H^{\rm NSM}$ and $A^{\rm NSM}$ respectively. The right diagram (b) appears via $d \geq 5$ interactions (indicated by the solid black disc) of pairs of $\chi$'s with the doublet-like Higgs states arising when integrating out a heavy Dirac fermion $SU(2)$-doublet.}
		\label{fig:DMSM_int}
	\end{center}
\end{figure}

In the absence of co-annihilation, the thermally averaged annihilation cross section for a pair of DM particles at temperature $T$ can be expanded as
\begin{equation}
	\langle\sigma_{\chi\chi} v \rangle \equiv \left\langle \sigma\left(\chi\chi\to {\rm SM}\right) v \right\rangle = a + b \langle v^2 \rangle +\mathcal{O}(\langle v^4 \rangle) = a+\frac{6 b}{x}+\mathcal{O}(\frac{1}{x^2}) \;,
\end{equation}
where $x = m_\chi/T$. After integrating over the thermal history of the Universe until the freeze-out temperature $T_F = m_\chi/x_F$, the thermal relic density is obtained
\begin{equation}
\Omega h^2 = 0.12\left(\frac{80}{g_*}\right)^{1/2}\left(\frac{x_F}{25}\right) \left( \frac{2.3\times 10^{-26} \mathrm{cm^3/s}}{\langle \sigma v\rangle_{x_F}}\right)\;,\qquad \langle \sigma v\rangle_{x_F} \equiv a+ \frac{3b}{x_F}\;.
\end{equation}

The interactions of the singlet fermion $\chi$ with SM particles depicted in Fig.~\ref{fig:DMSM_int} arise via the couplings to the extended Higgs basis states given in Eqs.\eqref{eq:gxxHSM1}--\eqref{eq:gxxAS} and the mixing of extended Higgs basis states into mass eigenstates,
\begin{equation} \label{eq:gxxhi}
	g_{\chi\chi h_i} = S_{h_i}^{\rm SM} g_{\chi\chi H^{\rm SM}} + S_{h_i}^{\rm NSM} g_{\chi\chi H^{\rm NSM}} + S_{h_i}^S g_{\chi\chi H^S}\;, \qquad g_{\chi\chi a_i} = P_{a_i}^{\rm NSM} g_{\chi\chi A^{\rm NSM}} + P_{a_i}^S g_{\chi\chi A^S} \;. 
\end{equation}
The singlet states $H^S$ and $A^S$ do not couple to SM particles, thus, assuming a type II 2HDM Yukawa structure, the couplings of the mass eigenstates to up-type quarks are given by
\begin{equation}
	g_{u h_i} = \left( S_{h_i}^{\rm SM} + \frac{S_{h_i}^{\rm NSM}}{\tan\beta} \right) \frac{m_u}{\sqrt{2} v}\;, \qquad g_{u a_i} = i \frac{P_{a_i}^{\rm NSM}}{\tan \beta} \, \frac{m_u}{\sqrt{2} v} \;, \label{eq:guhi}
\end{equation}
and to down-type quarks by
\begin{equation}
	g_{d h_i} = \left( S_{h_i}^{\rm SM} - S_{h_i}^{\rm NSM} \tan\beta \right) \frac{m_d}{\sqrt{2} v}\;, \qquad g_{d a_i} = i P_{a_i}^{\rm NSM} \tan \beta \, \frac{m_d}{\sqrt{2} v}\;. \label{eq:gdhi}
\end{equation} 
For completeness, we record the couplings to pairs of vector bosons
\begin{equation}
	g_{W^+ W^- h_i} = \frac{2 m_W^2}{v} S_{h_i}^{\rm SM} \;, \qquad g_{Z Z h_i} = \frac{m_Z^2}{v} S_{h_i}^{\rm SM} \;, \qquad g_{W^+ W^- a_i} = g_{Z Z a_i} = 0 ~.
\end{equation}

The contribution to $\langle \sigma v \rangle_{x_F}$ from annihilation into pairs of quarks ($\chi\chi \to q\bar{q}$) from the $s$-channel exchange of the CP-even Higgs bosons is given by
\begin{equation}\label{eq:relicEven}
	\langle \sigma v \rangle_{x_F}^{q\bar{q},p} =\frac{3}{2\pi} \frac{3}{4} \, \frac{T}{m_\chi} \left( 1-\frac{m_q^2}{m_\chi^2} \right)^{3/2}\left|\sum_i \mathcal{A}_{h_i}^{q\bar{q}} \right|^2\;, \qquad \mathcal{A}_{h_i}^{q\bar{q}} = \frac{- g_{\chi\chi h_i} \, g_{q h_i} \, m_\chi}{ \left(m_{h_i}^2 - 4 m_\chi^2 \right)} \; ,
\end{equation}
and from the exchange of CP-odd Higgs boson by
\begin{equation}\label{eq:relicOdd}
	\langle \sigma v \rangle_{x_F}^{q\bar{q},s} =\frac{3}{2\pi} \, \left( 1-\frac{m_q^2}{m_\chi^2} \right)^{1/2} \left|\sum_i \mathcal{A}_{a_i}^{q\bar{q}} \right|^2 \;, \qquad \mathcal{A}_{a_i}^{q\bar{q}} = \frac{-g_{\chi \chi a_i} \, g_{q a_i} \, m_\chi}{ \left(m_{a_i}^2 - 4 m_\chi^2 \right)} \;,
\end{equation}
for $|m_{\Phi_i} - 2 m_\chi| \gg \Gamma_{\Phi_i}$ with $\Gamma_{\Phi_i}$ the width of the Higgs mass eigenstate $\Phi_i$. Note, that there is no interference between the contributions listed in Eqs.~(\ref{eq:relicEven}) and (\ref{eq:relicOdd}) since the scalar Higgs bosons exchange contribution is $p$-wave suppressed while the annihilation cross section via pseudoscalar Higgs bosons is $s$-wave. For typical freeze-out temperatures $m_\chi/T_F \simeq 25$, the contribution from CP-even Higgs bosons to $\langle \sigma v \rangle_{x_F}$ is suppressed by $3 T_F / 4 m_\chi \sim 1/30$ compared to the contribution from CP-odd Higgs bosons, as long as $m_q/m_\chi \ll 1$ such that the kinematic correction from the quark mass is irrelevant.

Besides via Higgs bosons, ($\chi\chi \to q\bar{q}$) annihilations can also be mediated by the $s$-channel exchange of $Z$ bosons. This is accounted for by extending the sum in Eq.~\eqref{eq:relicOdd} to include the $s$-wave amplitudes mediated by both the longitudinal polarization of the $Z$ boson, i.e. the (neutral) Goldstone mode $G^0$, 
\begin{equation} \label{eq:relicAmpG}
	\mathcal{A}_{G^0}^{q\bar{q}} = \frac{-g_{\chi \chi G^0} \, g_{q G^0} \, m_\chi}{ \left(m_Z^2 - 4 m_\chi^2 \right)}\; , 
\end{equation}
as well as the transversal polarizations of the $Z$ boson
\begin{equation}\label{eq:relicZ}
	\mathcal{A}_Z^{q\bar{q}} = - \frac{m_q}{m_\chi} \frac{g_{\chi\chi Z} \, g_{qZ} \, m_\chi}{\left( m_Z^2 - 4 m_\chi^2 \right)} \; .
\end{equation}
The couplings of the Goldstone mode to up-type and down-type quarks, respectively, are
\begin{equation} \label{eq:Goldcoup}
	g_{uG^0} = i \frac{m_u}{\sqrt{2} v}~, \qquad g_{dG^0} = -i \frac{m_d}{\sqrt{2} v}~,
\end{equation}
and the relevant axial-vector coupling of the transversal polarizations of the $Z$ boson to quarks are
\begin{equation}
	g_{uZ} = -g_{dZ} = \frac{g_1}{4 s_W}~.
\end{equation}
The coupling of the $Z$ boson to the Majorana fermion can be read off from Eqs.~\eqref{eq:EFTmu} or~\eqref{eq:EFTmuHbasis}
\begin{equation} \label{eq:gxxZ}
	g_{\chi\chi Z} = -\frac{v^2}{|\mu|^2} \alpha \frac{g_1}{2 s_W} \cos 2\beta ~.
\end{equation}
Note, that the $s$-wave contribution to the annihilation cross section from the transversal polarization of the $Z$ boson is suppressed with respect to that of its longitudinal polarization (the neutral Goldstone mode) by $\mathcal{A}_Z/\mathcal{A}_{G^0} = -(m_Z^2/ 4 m_\chi^2) \sim -0.023 \times (m_\chi/300\,{\rm GeV})^{-2}$. 

All the amplitudes appearing in Eqs.~\eqref{eq:relicEven}-\eqref{eq:relicZ} are proportional to the Yukawa couplings. Due to the hierarchy of the Yukawa couplings, the contribution to the thermal cross section from ($\chi\chi \to q\bar{q}$) annihilations will be dominated by the heaviest accessible quarks, i.e. top-quarks for $m_\chi > m_t \sim 173\,$GeV and bottom quarks for lighter $m_\chi$. In the latter case the $p$-wave contribution from the transversal polarization of the $Z$ may become relevant, since in contrast to its $s$-wave contribution listed above it is not chirality suppressed~(hence, not proportional to the Yukawa couplings). 

It is interesting to consider the size of the thermally averaged cross section obtainable via ($\chi\chi \to q\bar{q}$) annihilation. For example, if we assume $m_\chi > m_t$, such that $(\chi\chi \to t\bar{t})$ is kinematically allowed, and assume the dominant annihilation channel to be via the (neutral) Goldstone mode, for $m_t^2 \ll m_\chi^2$, Eqs.~\eqref{eq:relicOdd}, \eqref{eq:relicAmpG} and \eqref{eq:Goldcoup}, approximately lead to
\begin{equation} \label{eq:gxxGforOmega}
	\langle \sigma v \rangle_{x_F}^{q\bar{q}} \sim 2 \times 10^{-26}\,\frac{{\rm cm}^3}{{\rm s}} \left( \frac{\left|g_{\chi\chi G^0}\right|}{0.1} \right)^2 \left( \frac{m_\chi}{300\,{\rm GeV}}\right)^{-2}.
\end{equation}
Hence, the correct relic density $\Omega h^2 \sim 0.12$~\cite{Ade:2015xua} is obtained from ($\chi\chi \to q\bar{q}$) annihilations for couplings $|g_{\chi\chi G^0}| \sim 0.1$.

In addition to 	the ($\chi\chi \to q\bar{q}$) annihilation discussed above, there may be relevant contributions to $\left\langle \sigma v \right\rangle_{x_F}$ from ($\chi\chi \to \Phi_i \Phi_j$) annihilations, where $\Phi_i$ denotes a scalar or pseudoscalar Higgs mass eigenstate. Such processes can be mediated either via diagrams with a Higgs or a $Z$ boson in the $s$-channel, or via $t/u$-channel exchange of the Majorana fermion $\chi$ or the (heavy) $SU(2)$-doublet Dirac fermion we integrated out. In our EFT, the last possibility proceeds via the $\chi\chi\Phi_i \Phi_j$ contact interaction terms in Eq.~\eqref{eq:EFTmu}. Regardless of the type of diagram, the annihilation into a pair of CP-even ($\chi\chi \to h_i h_j$) or CP-odd Higgs bosons ($\chi\chi \to a_i a_j$) is $p$-wave suppressed, while ($\chi\chi \to h_i a_j$) annihilations are $s$-wave. The corresponding $s$-wave contribution to the thermally averaged annihilation cross section is given by
\begin{equation} \label{eq:sigvPhiPhi}
	\left\langle \sigma v \right\rangle_{x_F}^{ha} = \frac{1}{64 \pi m_\chi^2} \left\{ \left[1-\frac{\left(m_{h_i} + m_{a_j}\right)^2}{4 m_\chi^2}\right] \left[1-\frac{\left(m_{h_i} - m_{a_j}\right)^2}{4 m_\chi^2}\right] \right\}^{1/2} \left| \sum_{k} \mathcal{A}_{k}^{h_ia_j} \right|^2 \:,
\end{equation}
where the sum includes the $s$-wave amplitudes mediated by CP-odd scalars $\Phi_k = \{ a_S, A, G^0 \}$ in the $s$-channel
\begin{equation} \label{eq:sigvampAH_s_A}
\mathcal{A}_{\Phi_k}^{h_ia_j} = \frac{-2 m_\chi \, g_{\chi\chi \Phi_k} \, g_{h_i a_j \Phi_k}}{m_{\Phi_k}^2 - 4 m_\chi^2}\;,
\end{equation}
the amplitude mediated by transversally polarized $Z$ bosons in the $s$-channel
\begin{equation} \label{eq:sigvampAH_Z}
	\mathcal{A}_Z^{h_ia_j} = - g_{\chi\chi Z} \, \frac{g_1}{s_W} P_{a_j}^{\rm NSM} S_{h_i}^{\rm NSM} \frac{m_{h_i}^2 - m_{a_j}^2}{m_Z^2 - 4 m_\chi^2} \;,
\end{equation}
the amplitudes mediated by the Dirac fermion in the $t/u$-channel proceeding via contact interaction terms after integrating out the Dirac fermion $\Psi$
\begin{equation} \label{eq:sigvamp_t_higgsino}
	\mathcal{A}_\Psi^{h_ia_j} = -2 g_{\chi\chi h_i a_j} m_\chi \;,
\end{equation}
and the amplitude from the $t/u$-channel exchange of the Majorana fermion $\chi$
\begin{equation} \label{eq:sigvamp_t_chi}
	\mathcal{A}_\chi^{h_ia_j} = - 2 g_{\chi\chi h_i} \, g_{\chi\chi a_j} \left[ 1 + \frac{ 2 m_{a_j}^2}{ 4 m_\chi^2 - \left(m_{h_i}^2 + m_{a_j}^2\right) } \right] .
\end{equation}
In Eqs.~\eqref{eq:sigvampAH_s_A}--\eqref{eq:sigvamp_t_chi}, the $g_{\chi\chi\Phi_i}$ are the couplings of pairs of $\chi$'s to the Higgs mass eigenstates given in Eq.~\eqref{eq:gxxhi} [Eq.~\eqref{eq:gxxG0} and~\eqref{eq:gxxZ} for the coupling to the neutral Goldstone mode $g_{\chi\chi G^0}$ and the transversal polarizations of the $Z$ boson $g_{\chi\chi Z}$, respectively], the $g_{\Phi_i\Phi_j\Phi_k}$ ($g_{\Phi_i \Phi_j G^0}$) are the dimensionful trilinear couplings between different Higgs mass eigenstates (between the neutral Goldstone mode and two Higgs mass eigenstates), which are not related to the parameters of our EFT, but arise from the Higgs potential~(see Appendices of Ref.~\cite{Carena:2015moc} for details). The $g_{\chi\chi\Phi_i\Phi_j}$ are the ($\chi\chi\Phi_j\Phi_k$) couplings of dimension (mass)$^{-1}$ which can be read off from the Lagrangian Eqs.~\eqref{eq:EFTmu} or~\eqref{eq:EFTmuHbasis} taking into account the mixing of the Higgs mass eigenstates, Eqs.~\eqref{eq:hi},~\eqref{eq:ai}. 

After accounting for suppression of couplings arising from the requirement of an $m_h = 125\,$GeV SM-like Higgs mass eigenstate and from approximately satisfying the blindspot condition, the most relevant final states for the ($\chi\chi \to \Phi_i \Phi_j$) processes are ($\chi\chi \to a_S h$) and ($\chi\chi \to a_S h_S$). If kinematically accessible, they can compete with ($\chi\chi \to t\bar{t}$) annihilation. For both these channels, the amplitudes mediated by the singlet-like CP-odd $a_S$ in the $s$-channel may play an important role. However, their relevance to the total cross section is dictated by the coupling strengths $g_{h a_S a_S}$ and $g_{h_S a_S a_S}$ respectively, which as mentioned above are not related to our EFT parameters. Hence for simplicity, we will assume that these couplings are small, rendering these processes irrelevant for the relic density. 

Ignoring such Higgs exchange diagrams, the amplitude $\mathcal{A}_\Psi^{h_ia_j}$ mediated by a $t/u$-channel Dirac fermion, which we integrated out, is most relevant for the final state ($\chi\chi \to a_S h$). Ignoring the kinematic correction in Eq.~\eqref{eq:sigvPhiPhi} which is relevant only very close to threshold ($m_{h_i} + m_{a_i} = 2 m_\chi$), canonical values of the thermally averaged annihilation cross section $\left\langle \sigma v \right\rangle_{x_F} \sim 2 \times 10^{-26}\,{\rm cm}^3\,{\rm s}^{-1}$ can be achieved for 
\begin{equation}
	\left|g_{\chi\chi h a_S}\right| \approx \left|g_{\chi\chi H^{\rm SM} A^S} \right| \approx \left| - i \frac{v}{\mu} \left( \frac{\lambda \delta}{\mu} \sin 2\beta + \frac{2 \kappa \xi}{\mu^*} \right) \right| \sim 4 \times 10^{-4}\,{\rm GeV}^{-1}\;,
\end{equation}
where we have assumed the mixing of the CP-odd Higgs bosons to be small. This corresponds to couplings
\begin{equation} \label{eq:gxxhSforOmega}
	\delta \lambda \sin2\beta + 2 \kappa \xi \sim \left( \frac{\mu}{700\,{\rm GeV}} \right)^2~.
\end{equation}

For the channel ($\chi\chi \to a_S h_S$) the processes associated with a Majorana fermion $\chi$ in the $t/u$-channel can be relevant. Assuming again for simplicity that these processes dominate compared to the one associated to the interchange of a singlet pseudoscalar (i.e. $g_{h_S a_S a_S}$ is small), neglecting the threshold corrections for ($m_{h_S}+m_{a_S} \approx 2 m_\chi$), and corrections from singlet-doublet mixing (the latter potentially leading to ${\cal O}(1)$ suppression), the thermally averaged annihilation cross section $\left\langle \sigma v \right\rangle_{x_F} \sim 2 \times 10^{-26}\,{\rm cm}^3\,{\rm s}^{-1}$ can be achieved for a DM coupling to the singlets
\begin{equation} \label{eq:gxxSforOmega}
\left|g_{\chi\chi a_S}\right| \approx	\left|g_{\chi\chi h_S}\right| \sim 0.2 \left( \frac{m_\chi}{300\,{\rm GeV}}\right)^{1/2} .
\end{equation}
This implies
\begin{equation}
	\left|\kappa\right| \sim 0.15~\left( \frac{m_\chi}{300\,{\rm GeV}}\right)^{1/2}\;,
\end{equation}
where we assumed $v \ll \mu$ for the estimate on $\kappa$ such that $g_{\chi\chi h_S} \sim \sqrt{2} \kappa$.

Annihilations into pairs of vector bosons [$\chi\chi \to ZZ (W^+ W^-)$] do not play an important role for obtaining the thermal relic density as long as $m_\chi > m_t$. Final states consisting of two vector bosons with longitudinal polarizations are $p$-wave suppressed since they correspond to annihilations into a pair of CP-odd scalars (i.e. the neutral and charged Goldstone modes for the $Z$ and $W$ bosons, respectively). Annihilations into a pair of transversally polarized vector bosons or one transversally polarized and one longitudinally polarized vector boson are $s$-wave. However, such annihilations proceeding via $t/u$-channel exchange of the neutral (charged) components of the $SU(2)$-doublet fermion we integrated out correspond to $\chi\chi ZZ$ ($\chi\chi W^+ W^-$) contact interaction terms in our EFT which would only appear at dimension $d \geq 7$ and hence are strongly suppressed. For $m_\chi > m_t$, annihilations mediated by an $s$-channel Higgs or $Z$ boson are also suppressed: the coupling of the $s$-channel mediator to one transversally and one longitudinally (a pair of transversally) polarized vector bosons is proportional to the gauge couplings (squared). The couplings of the Higgs and Goldstone bosons to quarks, instead, are proportional to the top Yukawa coupling. The Higgs mediated channel is furthermore also $p$-wave suppressed.

To summarize, the proper value of the thermally averaged annihilation cross section $\left\langle \sigma v \right\rangle_{x_F} \sim 2 \times 10^{-26}\,{\rm cm}^3\,{\rm s}^{-1}$ leading to the observed relic density can be easily obtained when ($\chi \chi \to t\bar{t}$) annihilations are kinematically allowed, i.e. when $m_\chi > m_t$. The annihilation cross section will then typically be dominated by annihilations into top quarks mediated by the neutral Goldstone mode for which the proper value of $\left\langle \sigma v \right\rangle_{x_F}$ is achieved for $g_{\chi\chi G^0} \sim 0.1$, cf. Eq.~\eqref{eq:gxxGforOmega}. Annihilation into pairs of vector bosons is suppressed because it proceeds through smaller couplings. If kinematically allowed, the annihilation cross section into pairs of Higgs mass eigenstates may become large enough, cf. Eqs.~\eqref{eq:gxxhSforOmega},~\eqref{eq:gxxSforOmega}, although the annihilation into pairs of top quarks tends to be competitive or dominant unless the EFT parameters conspire to suppress the $g_{\chi\chi G^0}$ coupling.

For lighter DM candidates, $m_\chi < m_t$, achieving a sufficiently large annihilation cross section $\left\langle \sigma v \right\rangle_{x_F} \sim 2 \times 10^{-26}\,{\rm cm}^3\,{\rm s}^{-1}$ is more difficult. In this case, ($\chi\chi \to b\bar{b}$) annihilation is dominated by the $Z$ boson mediated $p$-wave contribution, and couplings are usually not sufficiently large to obtain the proper annihilation cross section. Annihilation into pairs of Higgs bosons requires large couplings between the different Higgs mass eigenstates in order to be sufficiently effective; in addition it is not easy to obtain a light enough Higgs mass spectrum ($m_{h_i} + m_{a_i} < 2 m_\chi$) while simultaneously satisfying collider constraints. Annihilation into pairs of vector bosons usually does not achieve sufficiently large cross sections either since they are controlled by gauge couplings. Hence, for $m_\chi < m_t$ avoiding over-closure of the Universe requires large couplings in the Higgs sector unless the annihilation cross sections via a Higgs boson $\Phi$ (a $Z$ boson) in the $s$-channel is boosted via resonant annihilation, $2 m_\chi \approx m_\Phi$ ($2 m_\chi \approx m_Z$). 

\subsection{EFT: Direct Detection}

Elastic $(\chi q - \chi q)$ scattering relevant for direct detection proceeds via the same diagrams as annihilation in the early Universe depicted in Fig.~\ref{fig:DMSM_int}, but interpreting them as $t$-channel exchanges of Higgs bosons. The exchange of CP-odd Higgs bosons leads to spin-dependent scattering, and the contribution of the Goldstone mode $G^0$ is furthermore suppressed by $q^2/m_Z^2$, where $q$ is the momentum transfer. In contrast, the exchange of CP-even Higgs bosons leads to spin-independent scattering. Since bounds on the spin-independent DM-nucleon scattering~(SIDD) cross section~\cite{Angloher:2015ewa,Agnese:2017jvy,Aprile:2017iyp,Cui:2017nnn} are much stronger than those on the spin-dependent scattering~(SDDD) cross section~\cite{Akerib:2017kat,Amole:2017dex} we focus on SIDD in the remainder of this section. 

Summing coherently over the contribution from all CP-even Higgs mass eigenstates, the $\chi$-proton SIDD cross section can be written as
\begin{equation} \label{eq:SIDD}
	\sigma_p^{\rm SI} = \frac{2 m_p^2}{\pi} \left( \frac{m_p m_\chi}{m_p + m_\chi} \right)^2 \left\{ \sum_{i=h,H,h_S} \left[ F_u \left(\frac{a_u}{m_u}\right)_i + F_d \left(\frac{a_d}{m_d}\right)_i \right] \right\}^2, 
\end{equation}
where $m_p$ is the mass of the proton, and the form factors (at zero momentum transfer) are\footnote{Note, that there are considerable uncertainties on the form factors. In this work, we use the default values used by \texttt{micrOMEGAs\_4.3.5} $\{ f^p_{Tu}, f^p_{Td}, f^p_{Ts} \} = \{ 0.0153, 0.0191, 0.0447\}$~\cite{Belanger:2013oya, Belanger:2010pz, Belanger:2008sj, Belanger:2006is}. While we are mainly interested in the DM--Higgs couplings which are not directly affected by the form factors, a different choice for the values of the form factors can be compensated by (small) redefinitions of other parameters.} $F_u = f_{Tu}^p + \frac{4}{27} \left( 1 - \sum_{q=u,d,s} f_{Tq}^p \right) \approx 0.15$ and $F_d = f_{Td}^p + f_{Ts}^p + \frac{2}{27} \left( 1 - \sum_{q=u,d,s} f_{Tq}^p \right) \approx 0.13$. The $\left( a_q/m_q\right)_i$ parametrize the contribution to the scattering amplitude from one Higgs mass eigenstate,
\begin{equation}
	\left( \frac{a_q}{m_q} \right)_i = - \frac{1}{\sqrt{2}} \frac{1}{m_{h_i}^2} \frac{g_{q h_i}}{m_q} g_{\chi \chi h_i} ~,
\end{equation}
where the $g_{\chi\chi h_i}$ and $g_{q h_i}$ are given in Eqs.~\eqref{eq:gxxhi} and~\eqref{eq:guhi}.

As discussed in Section~\ref{sec:EFT_Higgs}, the observation of the 125\,GeV Higgs boson at the Large Hadron Collider (LHC) with couplings close to that of a SM Higgs implies that our model must contain a CP-even Higgs eigenstate $h$ with $m_h \approx 125\,$GeV and composition $S_h^{\rm SM} \approx 1$, $\{S_h^{\rm NSM}, S_h^S \}\ll 1$. In addition, to avoid bounds on additional Higgs bosons from the LHC and the Large Electron-Positron Collider (LEP), the remaining CP-even mass eigenstates $H$ and $h_S$ must be either heavy $m_H \gg m_h$ or dominantly composed of $H^S$. The coupling of the $H^{\rm NSM}$ Higgs boson to down-type quarks is enhanced by $\tan\beta$, and therefore at large values of $\tan\beta$ the suppression induced by its large mass may be compensated by an enhancement of the coupling. This case allows for effective destructive interference between the $H^{\rm SM}$ and $H^{\rm NSM}$ contributions to the SIDD cross section~\cite{Huang:2014xua}. For values of $\tan\beta = {\cal O}(1)$, as we shall use in our work, the contribution of the non-standard Higgs bosons to the SIDD cross section will either be suppressed by $(a_q/m_q)^2 \propto 1/m_H^4 \ll 1/m_h^4$ or, in the case of mostly singlet states, by their small doublet components $(a_q/m_q)^2 \propto \{(S_{h_i}^{\rm SM})^2, (S_{h_i}^{\rm NSM})^2\} \leq (1-S_{h_i}^S)^2$. Under such conditions, the SIDD cross section will be dominated by the contribution from the SM-like state $h$. Hence, the current stringent bounds on the SIDD cross section lead to relevant constraints on $g_{\chi\chi h}$.

The bounds on $g_{\chi \chi h}$ may be estimated by considering the SIDD cross section, taking into account only the diagrams with an $h$ in the $t$-channel. We find from Eq.~\eqref{eq:SIDD}
\begin{equation} \begin{split}
	\left(\sigma^{\rm SI}_p \right)_h ~&= \frac{2 m_p^2}{\pi} \left( \frac{m_p m_\chi}{m_p + m_\chi} \right)^2 \left[ F_u \left(\frac{a_u}{m_u}\right)_h + F_d \left(\frac{a_d}{m_d}\right)_h \right]^2 \\
	&\sim 5 \times 10^{-9}\,{\rm pb}~\left(\frac{g_{\chi\chi h}}{0.1}\right)^2 \left( \frac{m_h}{125\,{\rm GeV}}\right)^{-4},
\end{split} \end{equation}
while the experimental limit is $\sigma^{\rm SI}_p(m_\chi = 300\,{\rm GeV}) \lesssim 3.3 \times 10^{-10}\,$pb~\cite{Cui:2017nnn}. Hence, values of $g_{\chi\chi h} \lesssim 0.025$ are necessary to fulfill these constraints. 
This range of values of $g_{\chi\chi h}$ may be compared with the values of $g_{\chi\chi G^0} \sim 0.1$ necessary to obtain the observed relic density as discussed in the previous section, cf. Eq.~\eqref{eq:gxxGforOmega}. In general, the couplings of $\chi$ to the Higgs mass eigenstates are expected to be of similar magnitude, or larger, than $g_{\chi\chi G^0}$, since they arise at the same order, or lower, in our EFT expansion. In particular, the coupling $g_{\chi \chi h}$ arises from dimension $d=5$ operators with the leading contribution suppressed by $v/\mu$, while $g_{\chi\chi G^0}$ arises via $d=6$ operators and is suppressed by $m_\chi v/|\mu|^2$. We therefore conclude that under the requirement of obtaining an acceptable relic density, the values of the coupling $g_{\chi\chi h}$ necessary to fulfill the constraints on the SIDD cross section may only be obtained if the blind spot condition, Eq.~\eqref{eq:bs}, is approximately fulfilled.

\subsection{Bounds on Couplings and Parameters of the EFT}

In the previous section we argued that in order to suppress the SIDD cross section below experimental limits, the model must sit close to the blind spot, Eq.~\eqref{eq:bs}. It is also possible that the amplitude mediated by the $h$ interferes destructively with those mediated by the other CP-even mass eigenstates $H$ and $h_S$, or both mechanisms may be at work simultaneously. As argued above, the amplitudes of the diagrams mediated by $H$ and $h_S$ are in general suppressed with respect to the amplitude of the $h$-mediated diagram, such that destructive interference will only be effective when the contribution from the SM-like Higgs $h$ is already suppressed by proximity to the blind spot condition. In order to demonstrate these properties, we shall briefly consider the simple case in which the singlet fields are heavy and play no role in the low energy effective theory. 

\begin{figure}
	\begin{center}
		\includegraphics[width=0.49\linewidth]{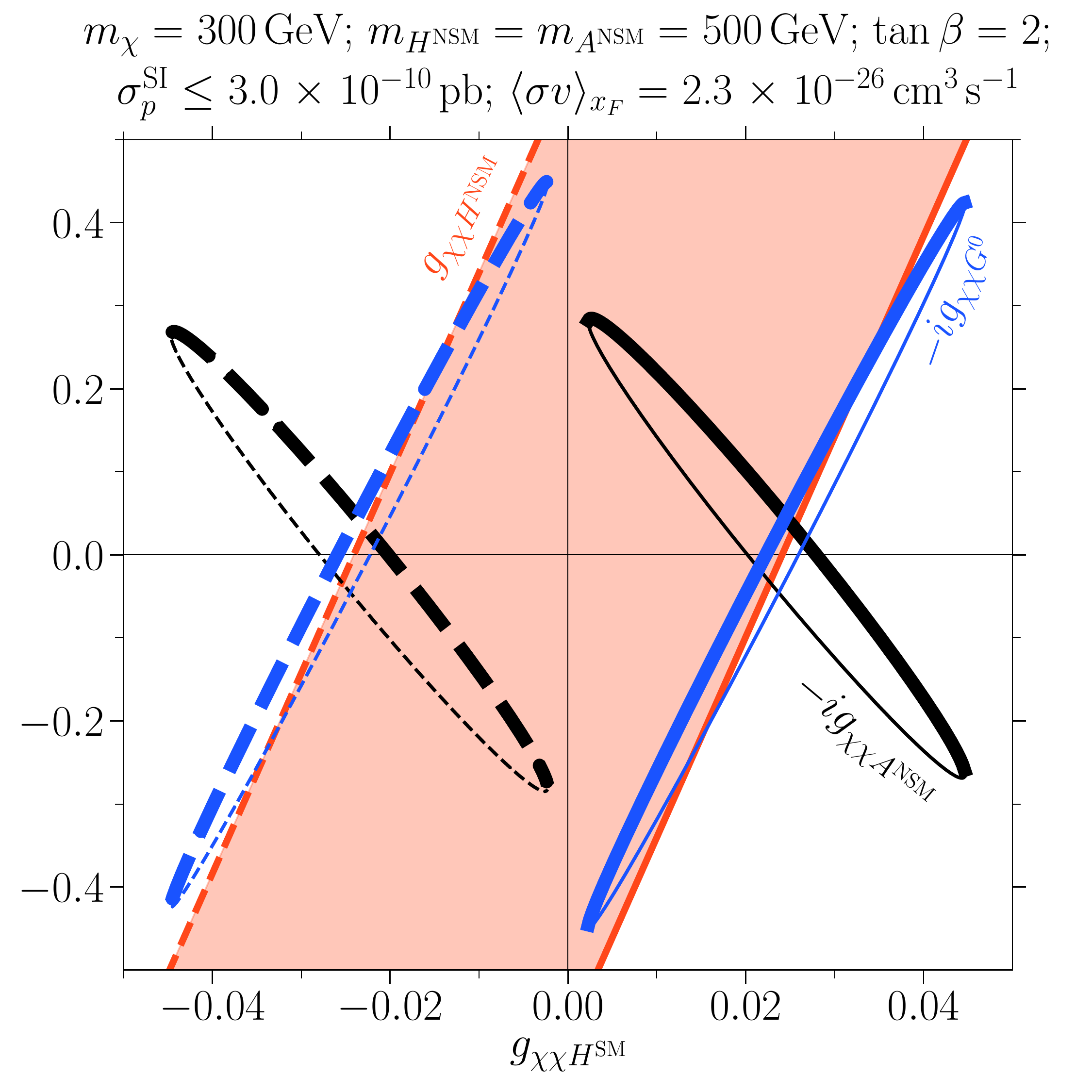}
		\includegraphics[width=0.49\linewidth]{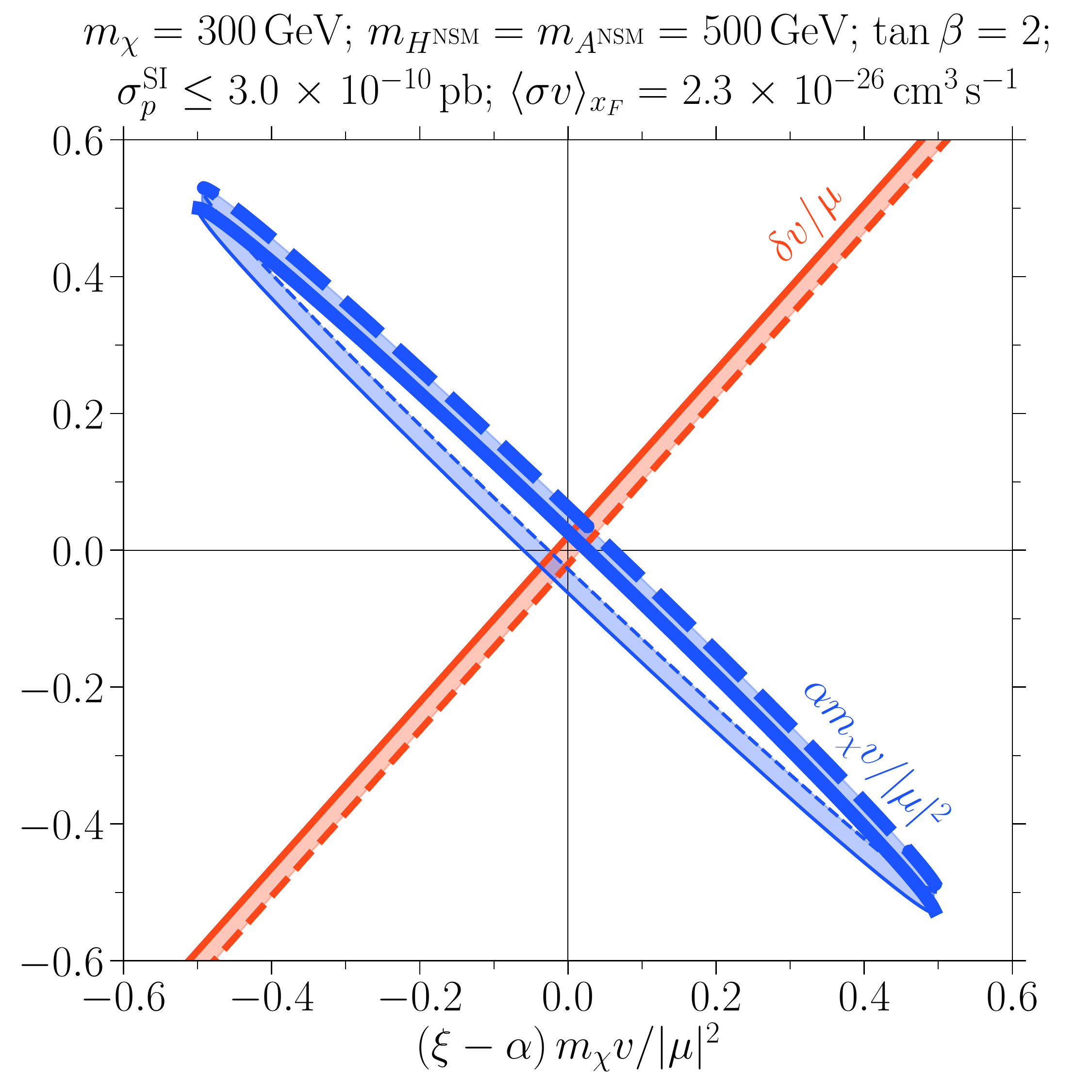}
		\caption{EFT parameters and couplings of DM to the CP-even and CP-odd Higgs bosons required to obtain the correct thermal relic density while concurrently satisfying SIDD constraints, for $\tan\beta = 2$, $m_\chi = 300\,$GeV, $m_{H^{\rm NSM}} = m_{A^{\rm NSM}} = 500\,$GeV, and decoupled singlet states. \textit{Left:} The orange shaded-region bounded by solid and dashed lines represents the CP-even Higgs bosons couplings consistent with the SIDD bounds, while the blue and black ellipses denote the couplings of DM to the (neutral) Goldstone mode and the heavy CP-odd Higgs boson yielding $\Omega h^2\sim0.12$ for CP-even Higgs couplings denoted by the corresponding solid or dashed lines, with thick and thin lines denoting two different solutions. \textit{Right:} Values of the EFT parameters consistent with the couplings shown in the left panel. Dashed and solid lines, as well as the shaded areas shown in this panel are in one-to-one correspondence with similar 
		lines and areas shown in the left panel.
		}
		\label{fig:DMSM_coup}
	\end{center}
\end{figure}

Fig.~\ref{fig:DMSM_coup} shows a representative example, illustrating the bounds on the parameters of our EFT and the couplings of DM to the CP-even and CP-odd Higgs bosons necessary to satisfy the SIDD experimental constraints and obtain the proper relic density concurrently. We used characteristic values of the DM mass, $m_{\chi} = 300\,$GeV, and the non-standard Higgs boson masses $m_{H^{\rm NSM}} = m_{A^{\rm NSM}}=500\,$GeV, and moderate values of $\tan\beta=2$. The singlets are decoupled $\{m_{H^S}, m_{A^S}\} \gg 500\,$GeV and perfect alignment is assumed in the doublet sector such that the Higgs basis states coincide with mass eigenstates. In the left panel we show the couplings of the DM to the non-standard Higgs bosons as a function of the DM coupling to the SM-like Higgs boson. As explained in the last section, ignoring the other CP-even Higgs bosons contributions, the coupling to the SM-like Higgs boson needs to be suppressed to satisfy the SIDD constraints, $g_{\chi\chi h} \lesssim 0.025$. As can be seen in Fig~\ref{fig:DMSM_coup}, this bound may be slightly relaxed in the presence of destructive interference with other Higgs bosons. The orange shaded region denotes the values of the couplings of DM to the CP-even Higgs bosons consistent with the SIDD bounds, with the boundaries denoting the values of these couplings saturating the current SIDD bound $\sigma_p^{\rm SI} = 3\times 10^{-10}\,$pb~(dashed and solid lines). The couplings of the CP-odd Higgs bosons are constrained only by the demand of obtaining a proper relic density.\footnote{Note, that the masses of the Higgs bosons are chosen such that ($\chi\chi \to h A$) annihilation are kinematically forbidden, hence, the relic density is set by ($\chi\chi \to t\bar{t}$) annihilation mediated dominantly by $G^0$ and $A^{\rm NSM}$.} The solid and dashed ellipses show the values of $g_{\chi\chi A^{\rm NSM}}$~(black) and $g_ {\chi\chi G^0}$~(blue) needed to obtain $\Omega h^2 \sim 0.12$, consistent with the SM-like Higgs couplings and the boundary values of the heavy CP-even Higgs couplings denoted by the solid and dashed lines, respectively. Thick and thin lines represent two different solutions for the CP-odd/Goldstone couplings for each set of values of the CP-even Higgs bosons couplings, with points in each of the ellipses in one-to-one correspondence with similar points in the other ellipses. The clear correlation between the couplings to the Goldstone mode and to the heavy CP-odd Higgs boson may be understood from the fact that for the given Higgs and DM masses, the suppression associated with the propagator contributing to the annihilation cross section is about a factor 3 weaker for the heavy CP-odd Higgs boson than for the Goldstone mode of mass $m_Z$: $3 \times 1/\left(m_{A^{\rm NSM}}^2 - 4 m_\chi^2 \right) \sim 1/\left(m_Z^2 - 4 m_\chi^2 \right)$. For couplings of DM to the CP-even Higgs bosons in the shaded area, for which the SIDD cross section would be smaller than the experimental limit, the couplings of the CP-odd Higgs boson and the Goldstone mode would take intermediate values between the ones represented by the two ellipses. 

The right panel of Fig.~\ref{fig:DMSM_coup} shows the three independent combinations of the parameters of the EFT that play a role in the determination of the relic density and the direct detection constraints. The shaded regions, solid lines, and dashed lines are in one-to-one correspondence with those in the left panel, and the shown correlations may be easily understood from the dependence of the couplings on these parameters, Eqs.~(\ref{eq:gxxHSM1}),~(\ref{eq:gxxHNSM}),~(\ref{eq:gxxG0}) and~(\ref{eq:gxxA}).

\section{NMSSM Ultraviolet Completion} \label{sec:NMSSM}
The EFT model discussed in the previous section extends the SM particle content by an additional $SU(2)$-doublet Higgs field, as well as a complex scalar field and a Majorana fermion both of which transform as singlets under the SM gauge group. There is also an additional $SU(2)$-doublet Dirac fermion which is assumed to be heavy and integrated out, yielding interactions of the Majorana fermion with the Higgs doublets. This particle content is very similar to the Higgs and neutralino sector of the NMSSM. Hence, decoupling the gluinos as well as the SUSY partners of the SM fermions from the theory, the NMSSM can serve as an ultraviolet completion for the EFT, preserving all the essential features discussed above. In addition, a range of mature numerical tools are available for the NMSSM which allows us to compute particle spectra and couplings and subsequently study the DM and collider phenomenology of the model. As we shall see, the NMSSM also contains a second region of parameter space where the DM candidate does not transform under the $Z_3$ symmetry, however, this region can nonetheless quite simply be mapped onto our EFT. In the SUSY case such a $Z_3$ symmetry is defined so that all chiral superfields transform by $e^{2\pi i/3}$, while gauge superfields transform trivially.

Reviews of the NMSSM can be found in Refs.~\cite{Ellwanger:2009dp,Maniatis:2009re}. The LHC phenomenology of the NMSSM Higgs and neutralino sectors has recently been investigated in Refs.~\cite{Gherghetta:2012gb,Christensen:2013dra,Cheung:2014lqa,Dutta:2014hma,Carena:2015moc,Cao:2016nix,Ellwanger:2016sur,Costa:2015llh,Baum:2017gbj,Ellwanger:2017skc,Kang:2013rj,King:2014xwa,Ellwanger:2015uaz}, and studies of the DM phenomenology include Refs.~\cite{Cheung:2014lqa,Cao:2015loa,Badziak:2015exr,Ellwanger:2016sur,Cao:2016nix,Cao:2016cnv,Beskidt:2017xsd,Badziak:2017uto}. The superpotential of the $Z_3$-invariant NMSSM contains no dimensionful parameters and is given by~\cite{Ellwanger:2009dp}
\begin{equation} \label{eq:WNMSSM}
	W = \epsilon^{ij} \left[ \lambda \widehat{S} (\widehat{H}_u)_i (\widehat{H}_d)_j - Y_u \widehat{\bar{u}} \widehat{Q}_i (\widehat{H}_u)_j - Y_d \widehat{\bar{d}} \widehat{Q}_i (\widehat{H}_d)_j - Y_e \widehat{\bar{e}} \widehat{L}_i (\widehat{H}_d)_j \right] + \frac{\kappa}{3} \widehat{S}^3 \; ,
\end{equation}
where we have written the $SU(2)$ indices explicitly. The Yukawa couplings $Y_u$, $Y_d$, and $Y_e$ should be understood to be matrices, and the left-handed quark (lepton) doublets $\widehat{Q}$ ($\widehat{L}$) as well as the up-type $\widehat{\bar{u}}$ and down-type right-handed quarks (leptons) $\widehat{\bar{d}}$ ($\widehat{\bar{e}}$) as vectors in family space. $\widehat{H}_u$ and $\widehat{H}_d$ are the usual Higgs doublet supermultiplets, and $\widehat{S}$ is a chiral supermultiplet which transforms as a singlet under the SM gauge group. 

The Higgs sector of the NMSSM consists of three neutral CP-even Higgs bosons, the real components of $H_u$, $H_d$, and $S$, two CP-odd neutral Higgs bosons, the imaginary components of $S$ and $A^{\rm NSM} = \sqrt{2} \, {\rm Im} \left( \cos\beta H_u^0 + \sin_\beta H_d^0 \right)$, and one charged Higgs $H^\pm$, with $\tan\beta = v_u/v_d$ and $v_u$ ($v_d$) the vev of $H_u$ ($H_d$). The remaining components of the Higgs doublets make up the longitudinal components of the $W$ and $Z$ bosons after electroweak symmetry breaking. The Higgs sector is controlled by the parameters
\begin{equation}
	p_i = \{\lambda, \kappa, \tan\beta, \mu, A_\lambda, A_\kappa\},
\end{equation}
where $\lambda$ and $\kappa$ are the dimensionless couplings appearing in the superpotential Eq.~\eqref{eq:WNMSSM}, $\mu \equiv \lambda \langle S \rangle$ with $\langle S \rangle$ the vev of $S$, and $A_\lambda$ and $A_\kappa$ are the dimensionful trilinear soft SUSY breaking couplings. Assuming CP conservation, one can without loss of generality choose all parameters to be real, and furthermore $\lambda$ and $\tan\beta$ to be positive, while $\kappa$ and the dimensionful parameters $\mu$, $A_\lambda$ and $A_\kappa$ can have both signs~\cite{Ellwanger:2009dp}. 

The Higgs sector is analogous to that of our EFT model, and can be rotated to the extended Higgs basis using Eqs.~\eqref{eq:Hbasis1}--\eqref{eq:Hbasis-1}. Including the usual soft SUSY breaking and $F$- and $D$-terms~\cite{Ellwanger:2009dp}, and ignoring radiative corrections for presentational purposes, see e.g. Ref.~\cite{Carena:2015moc}, the symmetric squared mass matrix for the neutral CP-even Higgs bosons in the basis $\{H^{\rm SM}, H^{\rm NSM}, H^S\}$ is~\footnote{Note, that Ref.~\cite{Carena:2015moc} uses the parameter $\overline{M}_Z \equiv m_Z^2 - \lambda^2 v^2$ and the $v=246\,$GeV convention, while we use $v = 174\,$GeV.} 
\begin{equation}
{\footnotesize
	\mathcal{M}_S^2 = \begin{pmatrix} \left[ m_Z^2 c_{2\beta}^2 + \lambda^2 v^2 s_{2\beta}^2 \right] & \left[ -\left(m_Z^2 - \lambda^2 v^2\right)s_{2\beta} c_{2\beta} \right] & \left[ 2 \lambda v \mu \left(1-\frac{M_A^2}{4\mu^2}s_{2\beta}^2 - \frac{\kappa}{2\lambda} s_{2\beta}\right) \right] \\
	& \left[ M_A^2 + \left(m_Z^2 - \lambda^2v^2\right)s_{2\beta}^2 \right] & \left[ - \lambda v \mu c_{2\beta} \left(\frac{M_A^2}{2\mu^2}s_{2\beta} + \frac{\kappa}{\lambda}\right) \right] \\
	& & \left[ \frac{1}{2} \lambda^2 v^2 s_{2\beta} \left(\frac{M_A^2}{2\mu^2}s_{2\beta}-\frac{\kappa}{\lambda}\right) + \frac{\kappa \mu}{\lambda}\left(A_\kappa + \frac{4 \kappa \mu}{\lambda}\right) \right] \end{pmatrix}} ,
\end{equation}
where $c_\beta \equiv \cos\beta$, $s_\beta \equiv \sin\beta$, and we have introduced
\begin{equation}
	M_A^2 \equiv \frac{2 \mu}{\sin 2\beta} \left( A_\lambda + \frac{\kappa \mu}{\lambda} \right).
	\label{eq:MA2}
\end{equation}

In the basis $\{A^{\rm NSM}, A^S\}$, the symmetric tree-level squared mass matrix for the CP-odd neutral Higgs bosons is 
\begin{equation}
	\mathcal{M}_P^2 = \begin{pmatrix} M_A^2 & \left[ \frac{1}{\sqrt{2}} \lambda v \left(\frac{M_A^2}{2\mu}s_{2\beta} - \frac{3\kappa\mu}{\lambda}\right) \right]
	\\ \left[ \frac{1}{\sqrt{2}} \lambda v \left(\frac{M_A^2}{2\mu}s_{2\beta} - \frac{3\kappa\mu}{\lambda}\right) \right] \hspace{.5cm} & \left[ \frac{1}{2} \lambda^2 v^2 s_{2\beta} \left(\frac{M_A^2}{4\mu^2} s_{2\beta} + \frac{3\kappa}{2\lambda}\right) - \frac{3\kappa A_\kappa \mu}{\lambda} \right] \end{pmatrix}. \label{eq:MP}
\end{equation}
Radiative corrections may be relevant and are incorporated at the two-loop level in the numerical results obtained with \texttt{NMSSMTools}~\cite{NMSSMTools}.

As in the case of our EFT model, the NMSSM must contain a CP-even 125\,GeV mass eigenstate mostly composed out of $H^{\rm SM}$ to accommodate the SM-like Higgs boson observed at the LHC~\cite{Aad:2015zhl,Khachatryan:2016vau}. This can be achieved either by making the remaining CP-even states $H$ and $h_S$ heavy,\footnote{Note, that we use the same notation for the Higgs states as in Section~\ref{sec:EFT}.} $\mathcal{M}_{S,22}, \mathcal{M}_{S,33} \gg \mathcal{M}_{S,11}$, the {\it decoupling} limit, or by setting $\mathcal{M}_{S,12}^2 \approx \mathcal{M}_{S,13}^2 \approx 0$, the {\it alignment-without-decoupling} limit~\cite{Carena:2015moc,Baum:2017gbj}. Perfect alignment is achieved for~\cite{Carena:2015moc}
\begin{equation} \label{eq:align}
	\lambda^2 = \frac{m_h^2 - m_Z^2 \cos(2\beta)}{2 v^2 \sin^2\beta}\;, \qquad \frac{M_A^2}{\mu^2} = \frac{4}{\sin^2 2\beta}\left(1-\frac{\kappa}{2\lambda} \sin 2\beta \right),
\end{equation}
and in the alignment limit the mass of the SM-like Higgs mass eigenstate is given by
\begin{equation}
	m_h^2 = \mathcal{M}_{S,11}^2 = m_Z^2 c_{2\beta}^2 + \lambda^2 v^2 s_{2\beta}^2 + \Delta(m_h^{\tilde{t}})\;, \label{eq:mh2}
\end{equation}	
where $\Delta(m_h^{\tilde{t}})$ are radiative corrections that are common to the MSSM. Note, that with respect to the MSSM, one obtains an additional contribution $(\lambda v s_{2\beta})^2$ to $m_h^2$ which allows for a 125\,GeV SM-like Higgs boson mass without the need for large radiative corrections for moderate values of $\tan\beta \lesssim 3$ if $\lambda \sim 0.7$. Intriguingly, the first alignment condition in Eq.~\eqref{eq:align}, which suppresses the $H^{\rm SM}-H^{\rm NSM}$ mixing, is satisfied for the same values of $\lambda$, namely $0.6 \lesssim \lambda \lesssim 0.7$. Alignment with the singlet [the second condition in Eq.~\eqref{eq:align}] is also easily achieved by judicious choices of $M_A$ and $\mu$. 

The neutralino sector of the NMSSM consists of the superpartners of the neutral electroweak gauge bosons, the bino $\widetilde{B}$ and neutral wino $\widetilde{W}^3$, the neutral Higgsinos $\widetilde{H}_d^0$ and $\widetilde{H}_u^0$ belonging to the respective Higgs doublets, and the singlino $\widetilde{S}$, the fermionic component of $\widehat{S}$. In the basis $\{\widetilde{B}, \widetilde{W}^3, \widetilde{H}_d^0, \widetilde{H}_u^0, \widetilde{S}\}$, the symmetric tree level neutralino mass matrix is 
\begin{equation} \label{eq:neumassmatrix}
	\mathcal{M}_{\chi^0}= \begin{pmatrix} M_1 & 0 & -m_Z s_W c_\beta & m_Z s_W s_\beta & 0
	\\ & M_2 & m_Z c_W c_\beta & -m_Z c_W s_\beta & 0
	\\ & & 0 & -\mu & - \lambda v s_\beta
	\\ & & & 0 & - \lambda v c_\beta
	\\ & & & & 2\kappa \mu / \lambda \end{pmatrix}, 
\end{equation}
where $M_1$ and $M_2$ are the bino and wino soft SUSY breaking masses. The neutralino mass eigenstates are given in terms of the interaction eigenstates by
\begin{equation}
	\chi_i = N_{i1} \widetilde{B} + N_{12} \widetilde{W}^3 + N_{i3} \widetilde{H}_d^0 + N_{i4} \widetilde{H}_u^0 + N_{i5} \widetilde{S} ~,
\end{equation}
where the mixing angles $N_{ij}$ are obtained from the diagonalization of the neutralino mass matrix Eq.~\eqref{eq:neumassmatrix}. 
Decoupling some of the neutralinos allows us to write down simple approximations for various mixing angles of particular interest. For example, if we neglect the singlino and wino contributions, diagonalizing the neutralino mass matrix Eq.~\eqref{eq:neumassmatrix} yields~\cite{Cheung:2014lqa}
\begin{align}
	\frac{N_{13}}{N_{11}} &= \frac{g_1}{\sqrt{2}} \frac{v}{\mu} \left[ \frac{ s_\beta + \left(m_{\chi_1}/\mu\right) c_\beta }{1 - \left( m_{\chi_1}/\mu\right)^2} \right] ,
	\nonumber\\ 
	\frac{N_{14}}{N_{11}} &= - \frac{g_1}{\sqrt{2}} \frac{v}{\mu} \left[ \frac{ c_\beta + \left(m_{\chi_1}/\mu\right) s_\beta }{1 - \left(m_{\chi_1}/\mu\right)^2} \right] , \label{eq:coupbino}
	\\ 
	N_{11} &= \left( 1 + \frac{N_{13}^2}{N_{11}^2} + \frac{N_{14}^2}{N_{11}^2} \right)^{-\frac{1}{2}}, \nonumber 
\end{align}
while if we neglect the bino and the wino component, we find~\cite{Cheung:2014lqa}
\begin{align}
	\frac{N_{13}}{N_{15}} &= \lambda \frac{v}{\mu} \left[ \frac{\left(m_{\chi_1}/\mu\right) s_\beta - c_\beta}{1-\left(m_{\chi_1}/\mu\right)^2} \right],
	\nonumber\\ 
	\frac{N_{14}}{N_{15}} &= \lambda \frac{v}{\mu} \left[ \frac{\left(m_{\chi_1}/\mu\right) c_\beta - s_\beta}{1-\left(m_{\chi_1}/\mu\right)^2} \right],
	\label{eq:coupsinglino} \\ 
	N_{15} &= \left( 1 + \frac{N_{13}^2}{N_{15}^2} + \frac{N_{14}^2}{N_{15}^2} \right)^{-\frac{1}{2}}. \nonumber	
\end{align}

In terms of the mixing angles, the couplings of the lightest neutralino to the Higgs basis states are
\begin{align} \label{eq:gxHang1}
	g_{\chi_1 \chi_1 H^{\rm SM}} &= \sqrt{2} \lambda N_{15} \left( N_{13} s_\beta + N_{14} c_\beta \right) + \left( g_1 N_{11} - g_2 N_{12} \right) \left( N_{13} c_\beta - N_{14} s_\beta \right) , 
	\\ g_{\chi_1 \chi_1 H^{\rm NSM}} &= \sqrt{2} \lambda N_{15} \left( N_{13} c_\beta - N_{14} s_\beta \right) - \left(g_1 N_{11} - g_2 N_{12} \right) \left( N_{13} s_\beta + N_{14} c_\beta \right) ,
	\\ g_{\chi_1 \chi_1 H^S} &= i g_{\chi_i \chi_j A^S} = \sqrt{2} \left[ \lambda N_{13} N_{14} - \kappa N_{15} N_{15} \right] ,
	\\ g_{\chi_1 \chi_1 A^{\rm NSM}} &= i \left[ \sqrt{2} \lambda N_{15} \left( N_{13} c_\beta + N_{14} s_\beta \right) - \left( g_1 N_{11} - g_2 N_{12} \right) \left( N_{13} s_\beta - N_{14} c_\beta \right) \right] ,
	\\ g_{\chi_1 \chi_1 G^0} &= i \left[ \sqrt{2} \lambda N_{15} \left( N_{13} s_\beta - N_{14} c_\beta \right) + \left( g_1 N_{11} - g_2 N_{12} \right) \left( N_{13} c_\beta + N_{14} s_\beta \right) \right] . \label{eq:gxHang-1}
\end{align}
From these, the couplings to the Higgs mass eigenstates can be obtained using Eq.~\eqref{eq:gxxhi} and the mixing angles of the Higgs mass eigenstates.

The singlino will play the role of the Majorana singlet in our EFT model. However, the NMSSM contains a second $SU(2)$-singlet neutralino, the bino. Unlike the singlino, the bino does not transform under the $Z_3$ symmetry since it stems from a gauge superfield. Besides allowing for an explicit mass term $(\frac{1}{2} M_1 \widetilde{B}\widetilde{B} + {\rm h.c.})$, this also results in different couplings to the Higgs doublets and the singlet than those found for the singlino. The region where the bino is the DM candidate can nonetheless be connected to our EFT in a straightforward way, as we will see in the following section.

\subsection{Top-down EFT} \label{sec:tdEFT}

In order to connect the NMSSM to our EFT from Section~\ref{sec:EFT}, we can construct a top-down EFT for the NMSSM Higgs and neutralino sector by integrating out the Higgsinos. This approach is valid as long as the Higgsino mass is large compared to the mass of the lightest neutralino. We will also assume the winos to be heavy, $M_2 \gg \{M_1, \mu\}$. Neglecting the Yukawa terms and ignoring the charged current interactions, the terms in the Lagrangian involving the neutral components of the Higgsinos are
\begin{equation} \begin{split} \label{eq:LtdEFT}
	\mathcal{L} ~&\supset (\widetilde{H}_u^0)^\dagger i \bar{\sigma}^\mu \left( \partial_\mu + i \frac{g_1}{2 \sin\theta_W} Z_\mu \right) \widetilde{H}_u^0 + (\widetilde{H}_d^0)^\dagger i \bar{\sigma}^\mu \left( \partial_\mu - i \frac{g_1}{2 \sin\theta_W} Z_\mu \right) \widetilde{H}_d^0 ~ \\
	& \qquad + \left\{ \lambda S \widetilde{H}_u^0 \widetilde{H}_d^0 + \left[ \lambda H_d^0 \widetilde{S} - \frac{g_1}{\sqrt{2}} (H_u^0)^\dagger \widetilde{B} \right] \widetilde{H}_u^0 + \left[ \lambda H_u^0 \widetilde{S} + \frac{g_1}{\sqrt{2}} (H_d^0)^\dagger \widetilde{B} \right] \widetilde{H}_d^0 + {\rm h.c.} \right\} .
\end{split} \end{equation} 
The corresponding equation of motion for $\widetilde{H}_u^0$ is
\begin{equation}
	0 = i \bar{\sigma}^\mu \left( \partial_\mu + i \frac{g_1}{2 \sin\theta_W} Z_\mu \right) \widetilde{H}_u^0 + \lambda S^\dagger (\widetilde{H}_d^0)^\dagger + \lambda (H_d^0)^\dagger \widetilde{S}^\dagger - \frac{g_1}{\sqrt{2}} H_u^0 \widetilde{B}^\dagger ~,
\end{equation} 
and for $\widetilde{H}_d^0$
\begin{equation} 
	0 = i \bar{\sigma}^\mu \left( \partial_\mu - i \frac{g_1}{2 \sin\theta_W} Z_\mu \right) \widetilde{H}_d^0 + \lambda S^\dagger (\widetilde{H}_u^0)^\dagger + \lambda (H_u^0)^\dagger \widetilde{S}^\dagger + \frac{g_1}{\sqrt{2}} H_d^0 (\widetilde{B})^\dagger ~.
\end{equation} 
Due to the Higgsino mass term $\lambda S \widetilde{H}_u^0 \widetilde{H}_d^0 \to \mu \widetilde{H}_u^0 \widetilde{H}_d^0$ when $S$ acquires a vev, and because of their identical couplings, we can interpret $\widetilde{H}_u^0$ as the right-handed and $\widetilde{H}_d^0$ as the left-handed component of a Dirac fermion with mass $\mu$. This allows us to use the equation of motion for $\widetilde{H}_u^0$ to integrate out $\widetilde{H}_d^0$ and vice versa. Keeping only terms leading to dimension $d\leq6$ operators when substituting the equations of motion into the Lagrangian, we obtain
\begin{equation} \begin{split}
	\widetilde{H}_d^0 = & - \frac{1}{\lambda^2 S^\dagger S} i \bar{\sigma}^\mu \left( \partial_\mu - i \frac{g_1}{2 \sin\theta_W} Z_\mu \right) \left[ \lambda (H_u^0)^\dagger \widetilde{S}^\dagger + \frac{g_1}{\sqrt{2}} (H_d^0) \widetilde{B}^\dagger \right] \\
	& \qquad - \frac{1}{\lambda S} \left[ \lambda H_d^0 \widetilde{S} - \frac{g_1}{\sqrt{2}} (H_u^0)^\dagger \widetilde{B} \right],
\end{split} \end{equation} 
and
\begin{equation} \begin{split}
	\widetilde{H}_u^0 =&~ - \frac{1}{\lambda^2 S^\dagger S} i \bar{\sigma}^\mu \left( \partial_\mu + i \frac{g_1}{2 \sin\theta_W} Z_\mu \right) \left[ \lambda (H_d^0)^\dagger \widetilde{S}^\dagger - \frac{g_1}{\sqrt{2}} H_u^0 \widetilde{B}^\dagger \right] \\
	& \qquad - \frac{1}{\lambda S} \left[ \lambda H_u^0 \widetilde{S} + \frac{g_1}{\sqrt{2}} (H_d^0)^\dagger \widetilde{B} \right].
\end{split} \end{equation} 
Substituting these into Eq.~\eqref{eq:LtdEFT} and keeping terms of dimension $d \leq 6$ (the same order used for our generic EFT in section~\ref{sec:EFT}), we find
\begin{equation} \begin{split} \label{eq:tdEFT}
	\mathcal{L} \supset &~ \frac{1}{\lambda S} \left\{ - \lambda^2 H_d^0 H_u^0 \widetilde{S} \widetilde{S} + \frac{\lambda g_1}{\sqrt{2}} \left[ (H_u^0)^\dagger H_u^0	- (H_d^0)^\dagger H_d^0 \right] \widetilde{B} \widetilde{S} + \frac{g_1^2}{2} (H_d^0)^\dagger (H_u^0)^\dagger \widetilde{B} \widetilde{B} \right\} + {\rm h.c.} \\
	& \qquad + \frac{1}{\lambda^2 S^\dagger S} \left[ \lambda (H_d^0)^\dagger \widetilde{S}^\dagger - \frac{g_1}{\sqrt{2}} H_u^0 \widetilde{B}^\dagger \right] i \bar{\sigma}^\mu \left( \partial_\mu - i \frac{g_1}{2 \sin\theta_W} Z_\mu \right) \left[ \lambda H_d^0 \widetilde{S} - \frac{g_1}{\sqrt{2}} (H_u^0)^\dagger \widetilde{B} \right] \\
	& \qquad + \frac{1}{\lambda^2 S^\dagger S} \left[ \lambda (H_u^0)^\dagger \widetilde{S}^\dagger + \frac{g_1}{\sqrt{2}} H_d^0 \widetilde{B}^\dagger \right] i \bar{\sigma}^\mu \left( \partial_\mu + i \frac{g_1}{2 \sin\theta_W} Z_\mu \right) \left[ \lambda H_u^0 \widetilde{S} + \frac{g_1}{\sqrt{2}} (H_d^0)^\dagger \widetilde{B} \right] \\
	& \qquad - \left( \kappa S \widetilde{S} \widetilde{S} + \frac{M_1}{2} \widetilde{B} \widetilde{B} + {\rm h.c.} \right) ,
	\end{split} \end{equation} 
where in the last line we have included the standard bino mass term and the singlet--singlino--singlino interaction term. Note, that because the singlino transforms under the $Z_3$ symmetry, it only gets a mass when the singlet $S$ acquires a vev. In contrast, the bino does not transform under the $Z_3$ and hence a soft SUSY breaking mass term $M_1$ is allowed.

Expanding $S$ around its vev $S \to \mu/\lambda + \hat{S}$ and correspondingly
\begin{equation}
	\frac{1}{\lambda S} \to \frac{1}{\mu} - \frac{\lambda \hat{S}}{\mu^2} + \mathcal{O}\left(\frac{\hat{S}^2}{\mu^3}\right), ~~~ \frac{1}{\lambda^2 S^\dagger S} \to \frac{1}{|\mu|^2} + \mathcal{O}\left(\frac{\hat{S}^2}{\mu^3}\right),
\end{equation} 
we can appreciate the similarities with the Lagrangian of our EFT model. In particular, the singlino has the same structure for the couplings as the Majorana fermion $\chi$ in Section~\ref{sec:EFT}, and we can map the couplings in Eq.~\eqref{eq:EFTmu} directly to those in the NMSSM via
\begin{equation} \label{eq:pmapS}
	\delta = -\alpha \to -\lambda^2~, \qquad \lambda \to \lambda~, \qquad \kappa \to \kappa~, \qquad \xi \to 0~. 
\end{equation}
The simple relation between the $\delta$ and $\alpha$ couplings $\delta = -\alpha$ arises because scalar and fermion components of chiral supermultiplets (vector and fermion components in the case of gauge supermultiplets) share the same couplings in SUSY models. The mapping above leads to the blind spot condition [cf. Eq.~\eqref{eq:bs}]
\begin{equation}
	\sin 2\beta = m_\chi/\mu~.
\end{equation}

In contrast to the singlino, the bino couples to different combinations of the Higgs doublets and the singlet. Such interactions would be obtained by writing down the EFT for the Higgs doublets and the singlet transforming under the $Z_3$, while assuming the Majorana fermion $\chi$ transforms trivially and has a Majorana mass $m_\chi = M_1$. In particular, comparing Eqs.~\eqref{eq:tdEFT} and~\eqref{eq:EFTmu} we see that the bino couples to $(H_d^0)^\dagger (H_u^0)^\dagger \widetilde{B} \widetilde{B}$ instead of $(H_d^0) (H_u^0) \chi\chi$, which can be compensated for by changing the sign of the coupling of the binos to the CP-odd states coming from the Higgs doublets, i.e. $A^{\rm NSM}$ and $G^0$. Keeping this in mind, we can map the couplings of the bino in Eq.~\eqref{eq:tdEFT} to those in the EFT, Eq.~\eqref{eq:EFTmu}, via
\begin{equation} \label{eq:pmapB}
	\delta = \alpha \to \frac{g_1^2}{2}~, \qquad \lambda \to \lambda~, \qquad \kappa = \xi \to 0~.
\end{equation}
The blind spot condition for the bino region is then
\begin{equation}
	\sin 2\beta = - m_\chi/\mu~.
\end{equation}

Note, that the bino couples with characteristic strength $g_1^2/2 \approx 0.06$ to Higgs doublet states, whereas the singlino couples to Higgs doublet states with characteristic strength $\lambda^2 \sim 0.4$, recalling that the presence of the 125\,GeV SM-like Higgs implies $\lambda \sim 0.6$.

The couplings of pairs of (on-shell) singlinos to the extended Higgs basis states and the (neutral) Goldstone mode can be directly read off from Eq.~\eqref{eq:tdEFT}, or from Eqs.~\eqref{eq:gxxHSM1},\eqref{eq:gxxHNSM}-\eqref{eq:gxxAS}, using the mapping of the parameters in Eq.~\eqref{eq:pmapS},
\begin{align} \label{eq:gSH1}
	g_{\widetilde{S}\widetilde{S} H^{\rm SM}} &= - \frac{\sqrt{2} v}{\mu} \lambda^2 \sin 2\beta + \frac{\sqrt{2} m_\chi v}{|\mu|^2} \lambda^2 ~, \\
	g_{\widetilde{S}\widetilde{S} H^{\rm NSM}} &= - \frac{\sqrt{2} v}{\mu} \lambda^2 \cos 2\beta ~, \\
	g_{\widetilde{S}\widetilde{S} H^S} = i g_{\widetilde{S}\widetilde{S} A^S} &= \frac{v^2}{\sqrt{2} \mu^2} \lambda^3 \sin 2\beta - \sqrt{2} \kappa \left( 1 - \frac{v^2}{|\mu|^2} \lambda^2 \right) ~, \label{eq:gSSS} \\
	g_{\widetilde{S}\widetilde{S} A^{\rm NSM}} &= -i\frac{\sqrt{2} v}{\mu} \lambda^2 + i \frac{\sqrt{2} m_\chi v}{|\mu|^2} \lambda^2 \sin 2\beta ~, \\
	g_{\widetilde{S}\widetilde{S} G^0} &= - i \frac{\sqrt{2} m_\chi v}{|\mu|^2} \lambda^2 \cos 2 \beta ~. \label{eq:gSH-1}
\end{align}
Similarly, the couplings of pairs of (on-shell) binos are given by
\begin{align} \label{eq:gBH1}
	g_{\widetilde{B}\widetilde{B} H^{\rm SM}} &= \frac{v}{\sqrt{2} \mu} g_1^2 \sin 2\beta + \frac{m_\chi v}{\sqrt{2} |\mu|^2} g_1^2 ~, \\
	g_{\widetilde{B}\widetilde{B} H^{\rm NSM}} &= \frac{v}{\sqrt{2} \mu} g_1^2 \cos 2\beta ~, \\
	g_{\widetilde{B}\widetilde{B} H^S} = i g_{\widetilde{B}\widetilde{B} A^S} &= - \frac{v^2}{2\sqrt{2} \mu^2} \lambda g_1^2 \sin 2\beta ~, \label{eq:gBBS} \\
	g_{\widetilde{B}\widetilde{B} A^{\rm NSM}} &= -i \frac{v}{\sqrt{2} \mu} g_1^2 - i \frac{m_\chi v}{\sqrt{2} |\mu|^2} g_1^2 \sin 2\beta ~, \\
	g_{\widetilde{B}\widetilde{B} G^0} &= i \frac{m_\chi v}{\sqrt{2} |\mu|^2} g_1^2 \cos 2\beta ~, \label{eq:gBH-1}
\end{align}
in agreement with the mapping of parameters in Eq.~\eqref{eq:pmapB}, keeping in mind the switch in the sign of the $(\chi\chi A^{\rm NSM})$ and $(\chi\chi G^0)$ couplings and the additional mass term $m_\chi = M_1$.

Note that Eq.~\eqref{eq:tdEFT} also induces bino--singlino--Higgs couplings which might be relevant for thermal production via co-annihilation when the bino and singlino are approximately mass degenerate, $M_1 \approx 2 \kappa \mu/\lambda$. For on-shell binos and singlinos, these couplings are
\begin{align}
	g_{\widetilde{B}\widetilde{S} H^{\rm SM}} &= - \frac{2 v}{\mu} \lambda g_1 \cos 2\beta ~, \\
	g_{\widetilde{B}\widetilde{S} H^{\rm NSM}} &= \frac{2 v}{\mu} \lambda g_1 \sin 2\beta + \frac{ \left( m_{\widetilde{B}} - m_{\widetilde{S}} \right) v}{|\mu|^2} \lambda g_1 ~, \\
	g_{\widetilde{B}\widetilde{S} H^S} = i g_{\widetilde{B}\widetilde{S} A^S} &= \frac{v^2}{\mu^2} \lambda^2 g_1 \cos 2\beta ~, \\
	g_{\widetilde{B}\widetilde{S} A^{\rm NSM}} &= i \frac{\left( m_{\widetilde{B}} + m_{\widetilde{S}} \right) v }{|\mu|^2} \lambda g_1 \cos 2\beta ~, \\
	g_{\widetilde{B}\widetilde{S} G^0} &= i \frac{ \left( m_{\widetilde{B}} + m_{\widetilde{S}} \right) v }{|\mu|^2} \lambda g_1 \sin 2\beta ~.
\end{align}

For mostly singlino-like (bino-like) DM with small Higgsino admixture, the couplings in Eqs.~\eqref{eq:gSH1}--\eqref{eq:gSH-1} [Eqs.~\eqref{eq:gBH1}--\eqref{eq:gBH-1}] give to good approximation the same numerical results as those in Eqs.~\eqref{eq:coupbino}--\eqref{eq:gxHang-1}. The differences between the two may be understood as originating from corrections due to higher dimensional operators in the EFT associated with the expansion of the denominator $\left[1-\left(m_\chi/\mu\right)^2 \right]$ in Eqs.~\eqref{eq:coupbino} and \eqref{eq:coupsinglino} in powers of $m_\chi^2/\mu^2$. Keeping only the first term, $1/\mu^2$, when replacing these expressions into Eqs.~\eqref{eq:gxHang1}--\eqref{eq:gxHang-1} is sufficient to reproduce the couplings obtained from the $d = 5$ and $d = 6$ operators, Eqs.~\eqref{eq:gSH1}--\eqref{eq:gBH-1}. 

\section{Dark Matter phenomenology} \label{sec:DMpheno}
\begin{table}
	\begin{center}
		\begin{tabular}{l@{\hspace{1cm}} c}
			\hline\hline
			$\tan\beta$ & $[1.5; 5]$ \\
			$\lambda$ & $ [0.5; 0.7]$ \\
			$\kappa$ & $[-0.3; +0.3]$ \\
			$\mu$ & $[-0.75; +0.75]$\,TeV \\
			$A_\lambda$ & $[-1.5; +1.5]$\,TeV \\
			$A_\kappa$ & $[-0.75; +0.75]$\,TeV \\
			$M_1$ & $[0; 1]$\,TeV \\
			\hline\hline
		\end{tabular}
		\caption{NMSSM parameter ranges used in \texttt{NMSSMTools} scan.}
		\label{tab:scanparams}
	\end{center}
\end{table}

Besides serving as an ultraviolet completion of our EFT, the NMSSM also provides a convenient computational basis since mature numerical tools are available for the analysis of both collider and DM phenomenology. In this section, we study the phenomenological properties of the NMSSM, going beyond the EFT validity conditions. Doing so, we can identify those properties that are shared with the EFT approach, while also determining the differences between the EFT and the full theory predictions. We use \texttt{NMSSMTools\_5.1.2}~\cite{NMSSMTools,Ellwanger:2004xm,Ellwanger:2005dv,Das:2011dg,Muhlleitner:2003vg} to compute NMSSM spectra and couplings and to subsequently test parameter points against a subset of the constraints implemented in \texttt{NMSSMTools} (see Ref.~\cite{NMSSMTools} for details). In particular, points are excluded if they feature an unphysical global minimum, soft Higgs masses much larger than the SUSY scale, and if the lightest neutralino is not the lightest SUSY particle. Furthermore, we require the spectrum to contain an $m_h \approx 125\,$GeV Higgs boson with couplings to SM particles compatible with those of the SM-like Higgs observed at the LHC. We also require compatibility with constraints from LEP, Tevatron, and the LHC on additional Higgs bosons and sparticles as implemented in \texttt{NMSSMTools}. For points passing these constraints we compute the relic density and the direct detection cross section with \texttt{micrOMEGAs\_4.3.5}~\cite{Belanger:2013oya, Belanger:2010pz, Belanger:2008sj, Belanger:2006is, Belanger:2005kh,Ellwanger:2006rn}. 

We perform a random scan over $10^9$ parameter points, drawing the parameters from linear-flat distributions over the ranges listed in Table~\ref{tab:scanparams}. The choice of parameter ranges is motivated by the phenomenology of a SM-like Higgs: for $\tan\beta \leq 5$, a 125\,GeV SM-like Higgs boson is obtained without the need for large radiative corrections to its mass, cf. Eq.~\eqref{eq:mh2}. The range for $A_\lambda$ is chosen to be larger than those for $\mu$ and $A_\kappa$ because approximate alignment implies, from Eqs.~\eqref{eq:MA2}~and~\eqref{eq:align},
\begin{equation}
A_\lambda = 2 \mu \left( \frac{1}{\sin 2\beta} - \frac{\kappa}{\lambda} \right).
\end{equation}
The range of the bino mass $M_1$ is chosen such that we obtain both the case where the lightest neutralino $\chi_1$ is mostly bino-like (i.e. when $M_1 < \{ \mu, 2\kappa\mu/\lambda \}$) and the case where the bino component of $\chi_1$ is negligible (i.e. when $\min(\mu, 2\kappa \mu/\lambda) \ll M_1$). In addition to the parameters listed in Table~\ref{tab:scanparams}, we scan over the stop mass $M_Q^3 = M_U^3$ in the range $[0.75; 2.5]\,$TeV and, for definiteness, we set the stop and sbottom mixing parameters $X_t \equiv (A_t - \mu \cot\beta) = 0$ and $X_b \equiv (A_b - \mu\tan\beta) = 0$,~\footnote{This choice minimizes radiative corrections to soft breaking parameters such as $A_\lambda$ and to the elements of the Higgs mass matrices, such that we can easily match analytical tree level results to numerical results from \texttt{NMSSMTools}. Including non-minimal third generation squark mixing would not affect our results, since we are only interested in the phenomenology of the Higgs and neutralino sectors for which the corresponding radiative corrections can be compensated for by shifts of the tree level parameters.} allowing for moderate radiative corrections to the Higgs masses. We decouple the remaining SUSY particles by setting all remaining sfermion mass parameters to 3\,TeV, the gluino mass parameter to $M_3 = 2\,$TeV, and, in order to minimize the wino component of the lightest neutralino, the wino mass parameter to $M_2 = 10\,$TeV. Note, that due to this choice of parameters we also satisfy all LHC bounds on sfermions and gluinos.

Due to our choice of parameters $M_2 \gg \{ M_1, |\mu|, |2\kappa\mu/\lambda| \}$, the heaviest neutralino $\chi_5$ will be wino like, while the wino component of the lighter mass eigenstates will be negligible. Furthermore, since we chose $|\kappa| \leq 0.3$, the singlino mass parameter $|2\kappa\mu/\lambda|$ will practically always be smaller than the Higgsino mass parameter $|\mu|$, while we chose the range of the bino mass parameter $M_1$ such that the bino can be both lighter and heavier than the singlino and the Higgsinos. This also ensures that the Higgsinos with mass parameter $\mu$ are always heavier than the lightest neutralino $\chi_1$, such that we omit the phenomenologically disfavored Higgsino DM region and we can map results onto our EFT, where $m_{\chi_1}/\mu$ appears as an expansion parameter. Hence, our DM candidate, the lightest neutralino, will be either mostly singlino-like or mostly bino-like with small Higgsino admixture facilitating Higgs mediated processes.

In all figures presented in this section, we compute the couplings of the neutralinos to Higgs mass eigenstates and of the SM particles to the Higgs mass eigenstates from the mixing angles $N_{ij}$ for the neutralinos and $S_{ij}$ ($P_{ij}$) for the CP-even (CP-odd) Higgs bosons as output from \texttt{NMSSMTools}. Computing the couplings from mixing angles, Eqs.~\eqref{eq:gxHang1}--\eqref{eq:gxHang-1}, takes into account sub-leading effects from the (small) admixture of additional neutralino components. These would only appear in the EFT through higher dimensional operators, or for wino effects from including operators arising from integrating out an $SU(2)$-triplet fermion in addition to the operators from integrating out an $SU(2)$-doublet fermion. Furthermore, using the output from \texttt{NMSSMTools} for the mixing angles also captures effects induced by the running of the parameters from the SUSY scale to the electroweak scale. 

\begin{figure}
	\begin{center}
		\includegraphics[width=.49\linewidth]{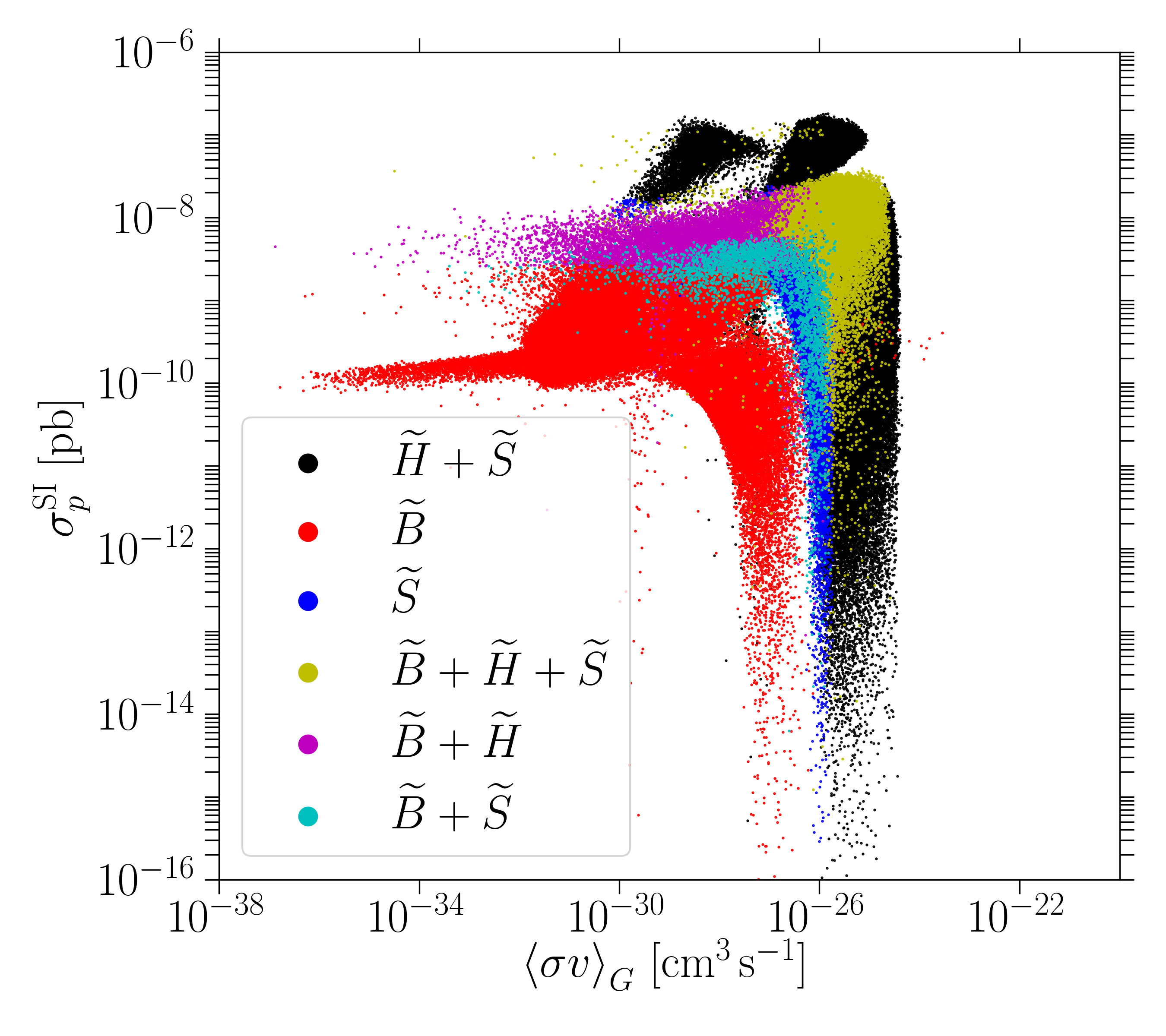}
		\includegraphics[width=.49\linewidth]{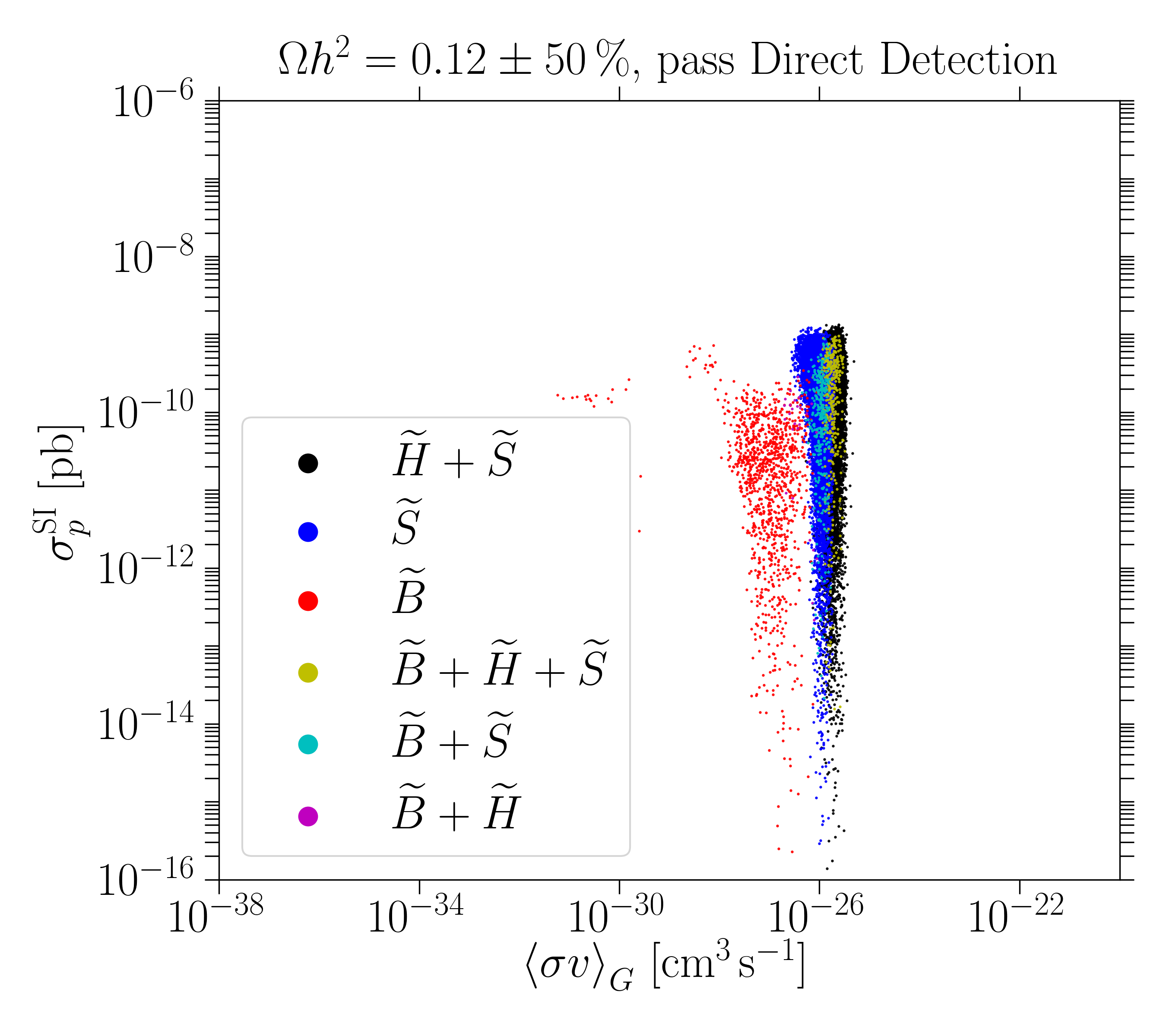}
		\caption{SIDD cross section $\sigma_p^{\rm SI}$ vs. the contribution of the longitudinal mode of the $Z$ boson (i.e. the Goldstone mode of the Higgs doublets) to the thermally averaged annihilation cross section as defined in Eq.\eqref{eq:relicOdd} with the amplitude given in Eq.~\eqref{eq:relicAmpG}. In the left panel, we show points from our parameter scan passing collider constraints, while points shown in the right panel are also required to have the correct relic density $\Omega h^2 = 0.12 \pm 50\,\%$ and satisfy bounds from direct detection experiments. The color coding indicates the compositions of the DM candidate as indicated in the legend. We denote points as purely bino $\widetilde{B}$ if $N_{11}^2 \geq 0.95$, Higgsino $\widetilde{H}$ if $N_{13}^2 + N_{14}^2 \geq 0.95$, or singlino $\widetilde{S}$ if $N_{15}^2 \geq 0.95$. Similarly, points are denoted as mixed if the sum of the square of corresponding mixing angles is $\sum N_{1i}^2 \geq 0.95$ but none of the individual contributions is sufficiently large to put them in one of the previous categories.}
		\label{fig:DD_sigv_singlinos}
	\end{center}
\end{figure}

In the NMSSM, apart from the processes discussed in detail in section~\ref{sec:EFT} , the annihilation cross section can be enhanced either by resonant annihilation or by co-annihilation~\cite{Griest:1990kh}. For the non-resonant annihilation cross sections, we checked numerically that for points with an acceptable relic density $\Omega h^2 \sim 0.12$, annihilations into top quarks typically make up for ${\cal O}(80\,\%)$ of the thermally averaged cross section $\left\langle \sigma v\right\rangle_{x_F}$, while annihilations into pairs of Higgs bosons usually account for ${\cal O}(20\,\%)$ of $\left\langle \sigma v\right\rangle_{x_F}$. This can be quiet generically understood to be due to the difficulty in obtaining a Higgs spectrum light enough to allow for the second final state, while evading Higgs phenomenology and collider constraints. Concentrating on ($\chi \chi \to t\bar{t}$), Fig.~\ref{fig:DD_sigv_singlinos} shows the SIDD cross section vs. the contribution to the thermal annihilation cross section from the diagrams mediated by the Goldstone mode obtained from Eq.~\eqref{eq:relicOdd} when taking into account only the Goldstone amplitude in Eq.~\eqref{eq:relicAmpG}. The left panel shows all points from our scan passing the Higgs and LHC constraints described above but before requiring the correct relic density or compatibility with direct detection limits. The dominant composition of the DM candidate is color coded and denoted as bino $\widetilde{B}$ if $N_{11}^2 \geq 0.95$~(red), singlino $\widetilde{S}$ if $N_{15}^2 \geq 0.95$~(blue). Similarly, points are denoted as mixed if the sum of the square of corresponding mixing angles is $\sum N_{1i}^2 \geq 0.95$ but none of the individual contributions is sufficiently large to put them in one of the previous categories. The behavior shown can be understood quite intuitively from our findings in sections~\ref{sec:EFT} and~\ref{sec:tdEFT}: first of all, we note that both the SIDD and the annihilation cross section for mostly bino DM are smaller than those for mostly singlino DM because the couplings to Higgs bosons are proportional to $g_1^2/2 \approx 0.06$ for binos while for singlinos the couplings are proportional to $\lambda^2 \sim 0.4$. Secondly, for all the different DM compositions, a striking feature of this plot is the relative independence of the SIDD cross section and $\left\langle \sigma v\right\rangle_{G}$, which can be suppressed independently from each other. The SIDD cross section is suppressed close to the blind spot conditions $m_\chi/\mu \to \pm \sin{2\beta}$. However, in this case the coupling to the Goldstone mode $g_{\chi\chi G^0} \propto m_\chi \cos{2\beta} / |\mu|^2$ remains sizable and thus the corresponding contribution to the annihilation cross section is not suppressed. Note, that the contributions to the thermal annihilation cross section from the Higgs mass eigenstates are usually smaller than those from the Goldstone mode: while the couplings are typically of the same order, the contributions from the CP-odd eigenstates are suppressed by
\begin{equation}
	\left(\frac{\mathcal{A}_{a_i}}{\mathcal{A}_{G^0}}\right) \sim \left( \frac{P_{a_i}^{\rm NSM}}{\tan\beta} \times \frac{m_{a_i}^2 - 4 m_\chi^2}{m_Z^2 - 4 m_\chi^2} \right)^2 \,,
\end{equation}
and the contributions from CP-even Higgs bosons are $p$-wave suppressed. As argued in section~\ref{sec:EFT_relic}, the contribution from the transverse polarizations of the $Z$ boson is generally also much smaller than the contribution from the Goldstone mode. 

\begin{figure}
	\begin{center}
		\includegraphics[width=.49\linewidth]{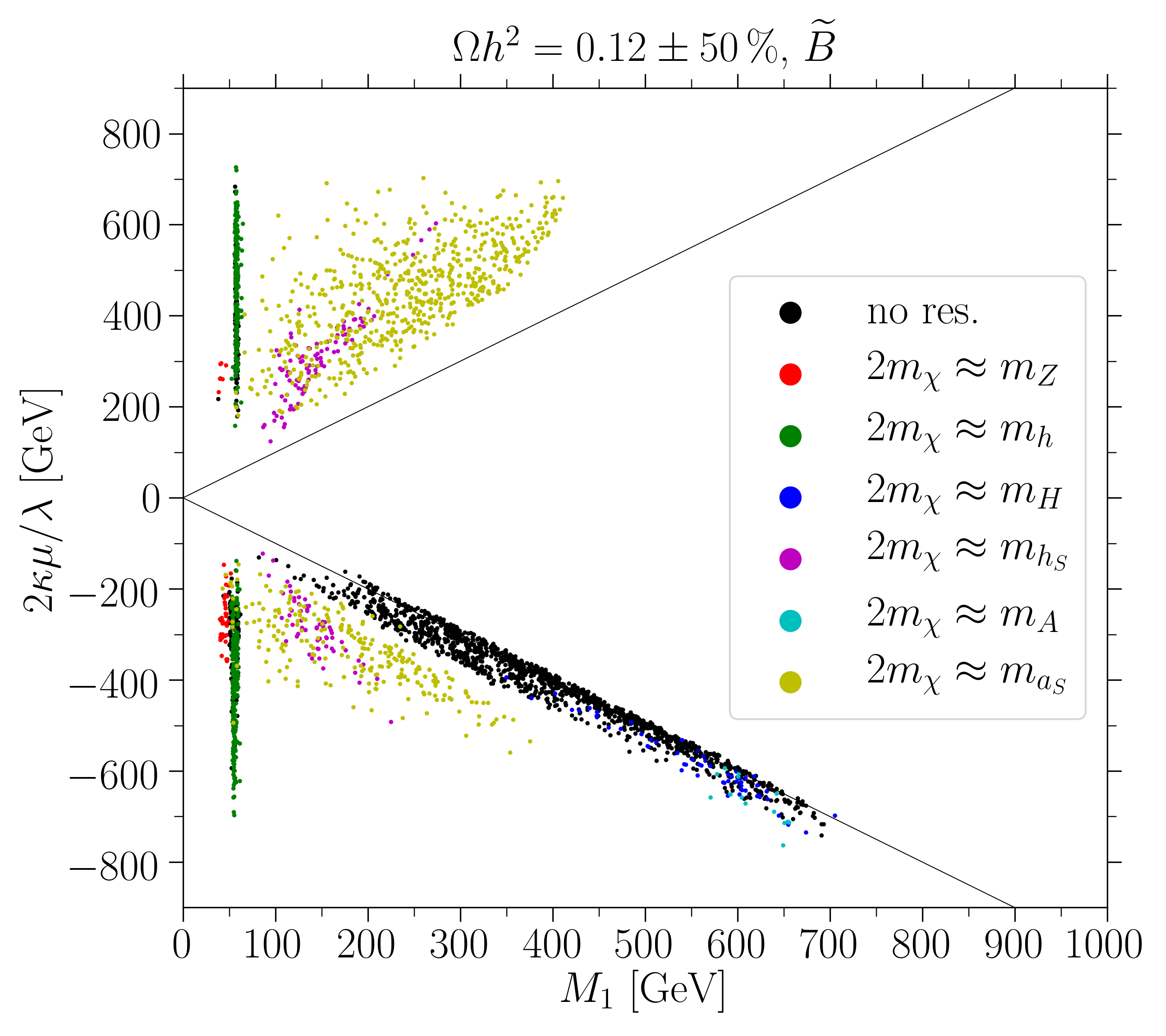}
		\includegraphics[width=.49\linewidth]{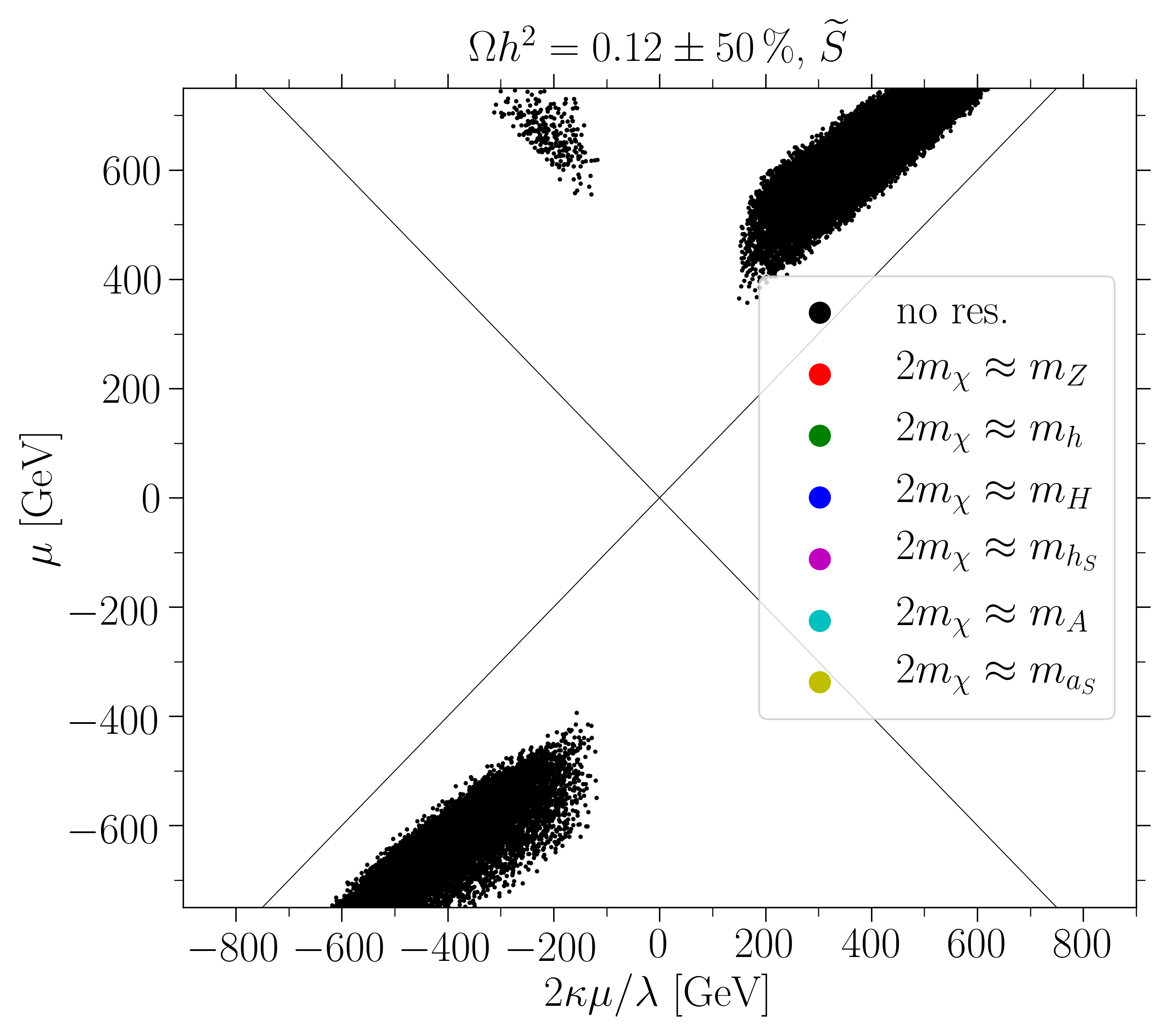}
		\caption{\textit{Left:} Points from our parameter scan with bino-like lightest neutralino and relic density $\Omega h^2 = 0.12 \pm 50\,\%$ in the bino mass ($M_1$) -- singlino mass ($2\kappa \mu/\lambda$) plane. \textit{Right:} Points with singlino-like lightest neutralino in the singlino mass ($2\kappa \mu/\lambda$) -- Higgsino mass ($\mu$) plane. For both panels, the color code indicates points where the lightest neutralino can pair-annihilate resonantly with the $s$-channel mediator with mass $\approx 2m_\chi$ as indicated in the legend.}
		\label{fig:DD_production}
	\end{center}
\end{figure}

The right panel of Fig.~\ref{fig:DD_sigv_singlinos} shows the points from our scan having approximately the correct relic density $\Omega h^2 = 0.12 \pm 50\,\%$ and satisfying bounds from SIDD and SDDD experiments~\cite{Aprile:2017iyp,Cui:2017nnn,Akerib:2017kat,Amole:2017dex}. For mostly singlino DM, the contribution from the Goldstone mode mediated amplitudes to the thermally averaged annihilation cross section is of the order of the canonical value $\langle \sigma v\rangle \sim 2 \times 10^{-26}\,{\rm cm}^2/{\rm s}$ yielding the observed relic density $\Omega h^2 \sim 0.12$~\cite{Ade:2015xua}~\footnote{Here and in the following we consider the relic density to be acceptable if if $\Omega h^2 = 0.12 \pm 50\,\%$. The width of this band is motivated by the strong sensitivity of the relic density calculation to the particle spectrum, and by the uncertainties in the calculation of particle masses~\cite{Bergeron:2017rdm}. Typical \texttt{NMSSMTools} uncertainties are of the order of a few GeV, as can for example be seen by comparing the results obtained by other spectrum generators~\cite{Goodsell:2014pla,Staub:2015aea}.}. In contrast, we find that for mostly bino DM the Goldstone mode does not mediate sufficiently large amplitudes to avoid over-closure of the Universe. Hence in the bino DM scenario, both resonant annihilation and co-annihilation play a very important role.

In Fig.~\ref{fig:DD_production} we show points from our numerical scan which have an acceptable relic density, $\Omega h^2 = 0.12 \pm 50\,\%$, with the color coding indicating the possibility of resonant annihilation. In the right panel of Fig.~\ref{fig:DD_production} we show points with mostly singlino DM candidate in the (singlino mass)--(Higgsino mass) plane, demonstrating that neither co-annihilation nor resonant annihilation is relevant for the singlino region~($M_1$ is always large in this region). In the left panel of Fig.~\ref{fig:DD_production} we show bino-like points in the (bino mass)--(singlino mass) plane. We find that points either resonantly annihilate via the $Z$ boson or one of the Higgs mass eigenstates, or, feature binos approximately mass degenerate with the singlino such that co-annihilation yields the correct relic density. In the latter case, it is in fact the annihilations of the mostly singlino like $m_{\chi_2}$ which set the relic density. We have thus found a new {\it well tempered} bino region: $m_{\chi_1}$ is mostly bino-like with very small couplings, evading direct detection constraints easily. However due to the presence of an almost mass degenerate singlino-like $m_{\chi_2}$~(which does not play a role in direct detection) which has significantly larger couplings, an observationally consistent relic density is easily obtained. The value of the $\mu$ parameter, and consequently the Higgsinos, tends to be about the same order as shown for the singlino-like DM in the right panel. 

\begin{figure}
	\begin{center}
		\includegraphics[width=.49\linewidth]{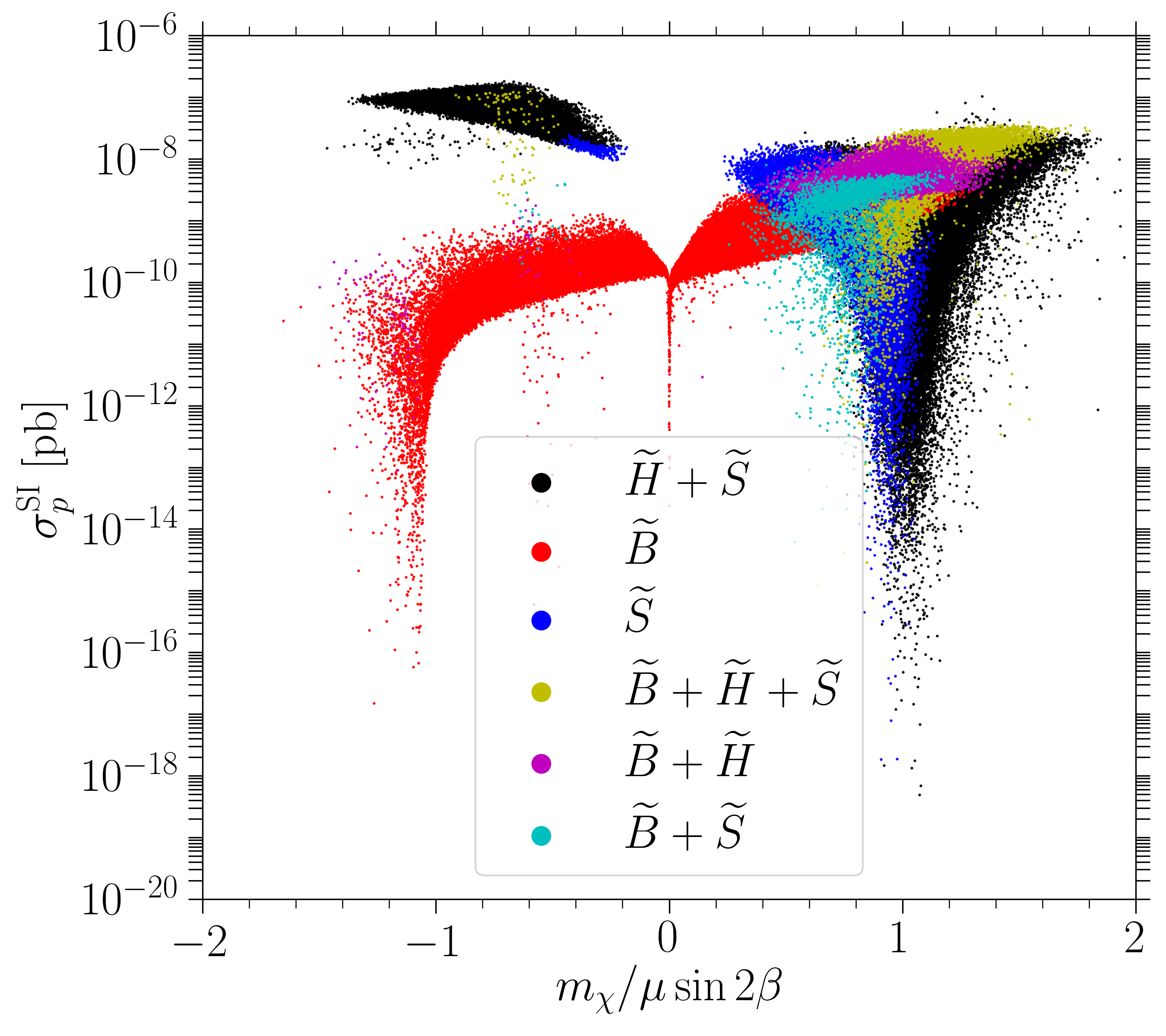}
		\includegraphics[width=.49\linewidth]{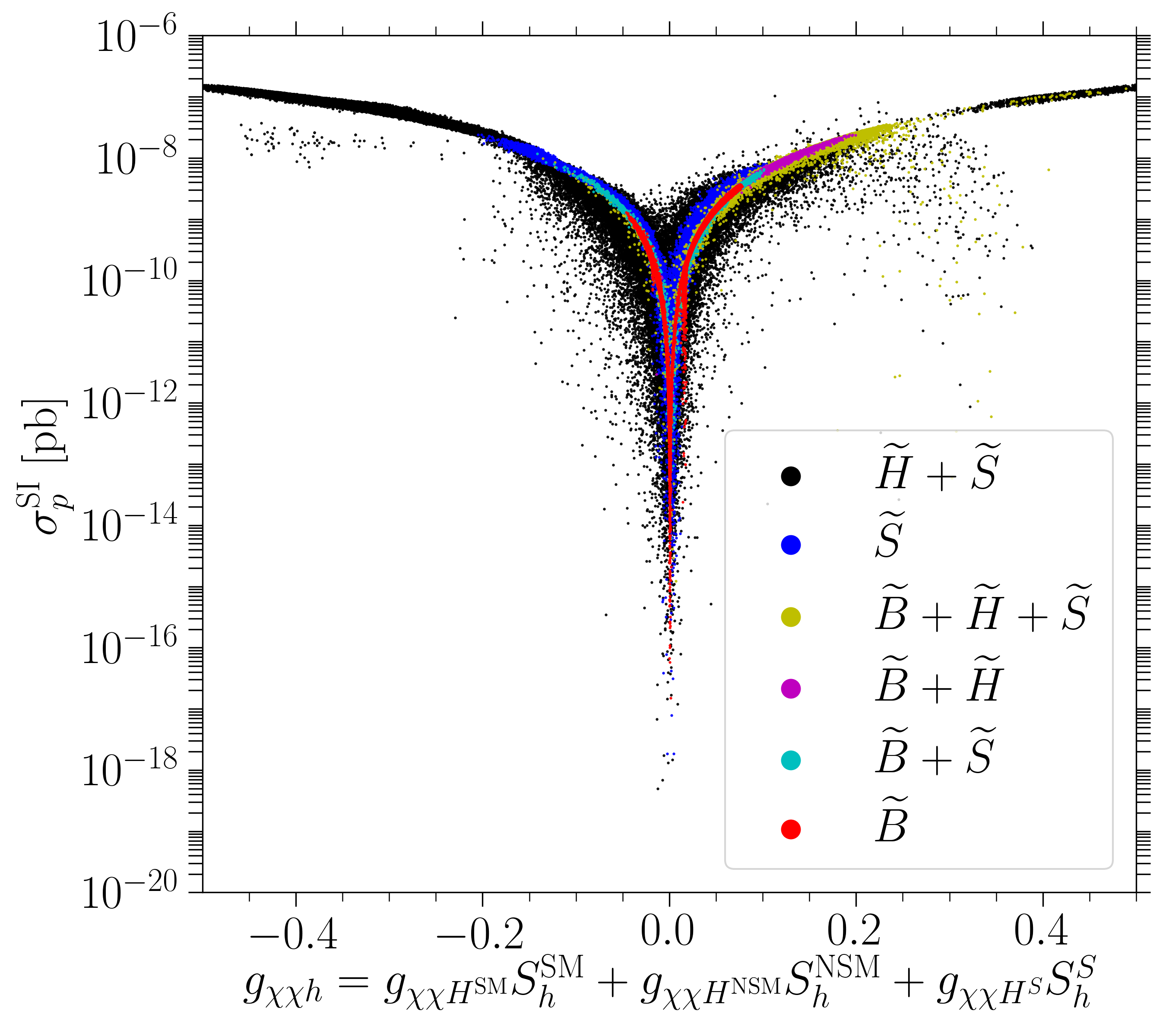}
		\caption{\textit{Left:} The SIDD cross section $\sigma_p^{\rm SI}$ for the points passing the required experimental collider constraints vs. $m_\chi/(\mu \sin 2\beta)$, where the blind-spot conditions are satisfied for $m_\chi/(\mu \sin 2\beta) = +1 (-1)$ for the singlino-Higgsino (bino-Higgsino) case. \textit{Right:} The spin independent cross section for the same points vs. the coupling of the DM candidate to the SM-like Higgs mass eigenstate. The color coding is the same as if Fig.~\ref{fig:DD_sigv_singlinos}.}
		\label{fig:DD_BS}
	\end{center}
\end{figure}

In the left panel of Fig.~\ref{fig:DD_BS} we show the SIDD cross section vs. $m_\chi/ ( \mu \sin 2\beta) $ for points from our scan passing the Higgs and collider constraints described above but before requiring the correct relic density or compatibility with direct detection limits. For bino DM, the SIDD cross section is suppressed when the blind spot condition $m_\chi/\mu = - \sin\beta$ is approximately satisfied, while for all other compositions of the DM candidate, usually singlino dominated, the SIDD cross section is suppressed for $m_\chi/\mu \approx \sin\beta$. Note that for bino DM candidates we also find suppression of the SIDD cross section for $m_\chi/\mu \approx 0$, which corresponds to $M_1 \ll \mu$ for which we do not find any parameter points with an acceptable relic density in our scan. This is because in this region neither co-annihilation with the singlino nor resonant annihilation with the Higgs bosons is possible, one of which would be required to boost the thermal annihilation cross section to avoid over-closure of the Universe.

In the right panel of Fig.~\ref{fig:DD_BS} we show the SIDD cross section vs. the DM coupling to the SM-like 125\,GeV Higgs mass eigenstate. Besides the coupling to the $H^{\rm SM}$ Higgs basis state, which vanishes at the respective blind spots, this takes into account the contributions to the coupling from the (small) admixtures of the $H^{\rm NSM}$ and $H^S$ interaction eigenstates to the 125\,GeV mass eigenstate. We find that for mostly bino points the SIDD cross section is very tightly correlated with the coupling to the 125\,GeV SM-like Higgs mass eigenstate, and SIDD cross sections satisfying the current experimental bounds can be achieved by suppression of the $g_{\widetilde{B}\widetilde{B} h}$ coupling. In the case of mostly singlino DM we find this correlation to be looser, indicating that the SIDD cross sections must be suppressed by additional mechanisms.

\begin{figure}
	\begin{center}
		\includegraphics[width=.49\linewidth]{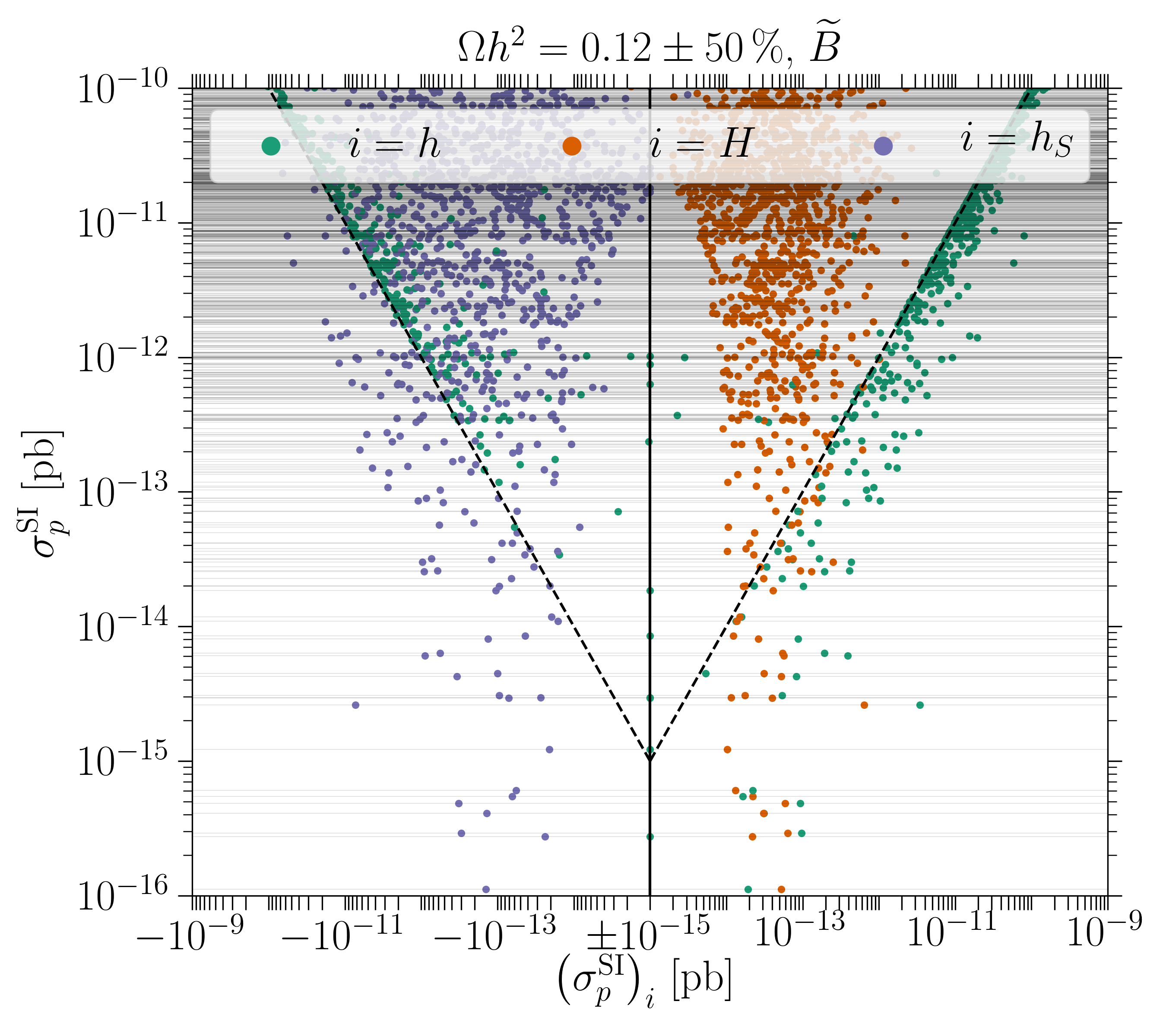}
		\includegraphics[width=.49\linewidth]{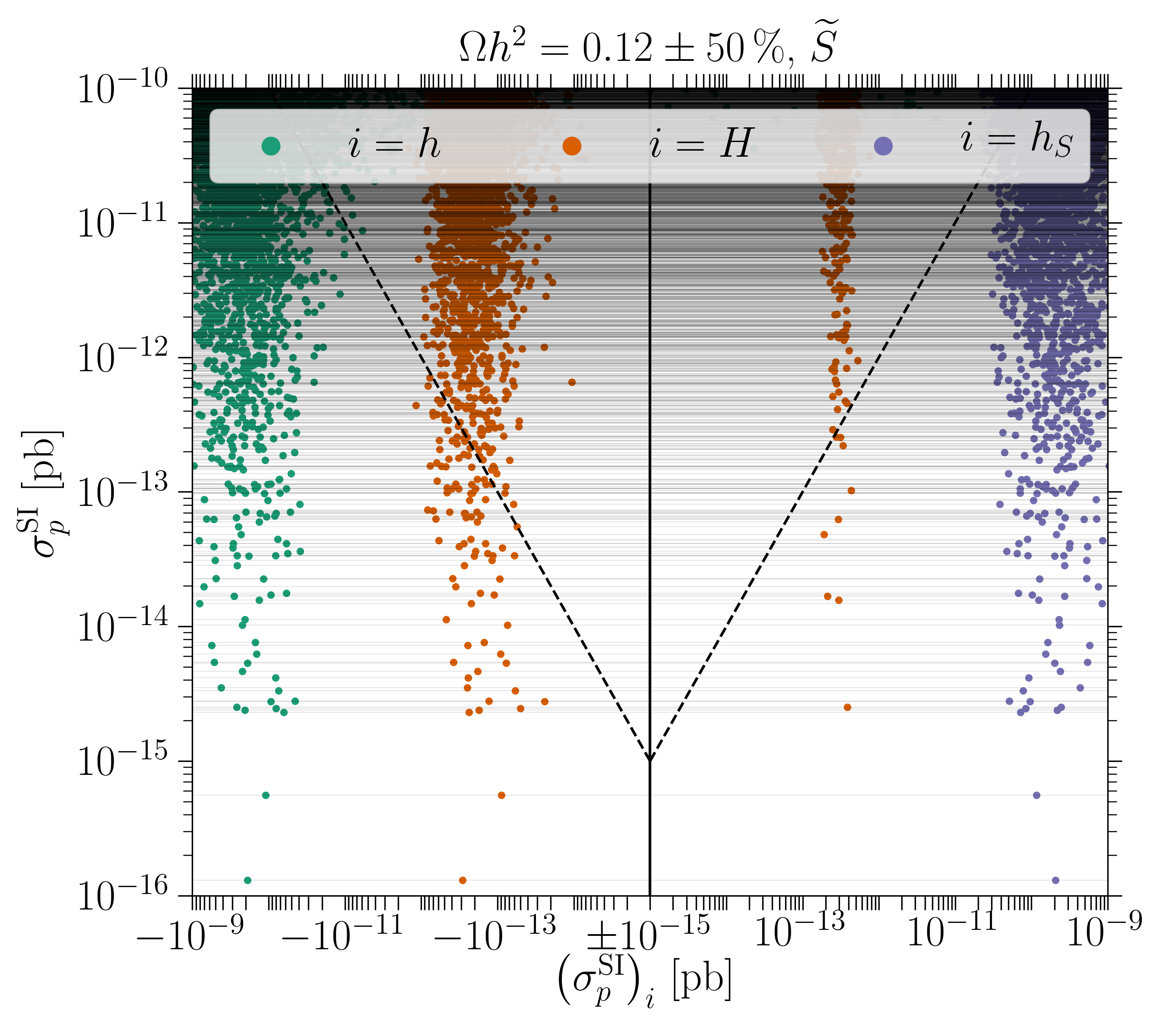}
		\caption{SIDD cross section $\sigma_p^{\rm SI}$ vs. the contribution $\left( \sigma_p^{\rm SI} \right)_i$ assuming only one CP-even Higgs mass eigenstate $h_i = \{ h, H, h_S \} $ multiplied by the sign of its amplitude, cf. Eq.~\eqref{eq:SIDD}. The dashed diagonal lines indicate $\left|\left( \sigma_p^{\rm SI} \right)_i\right| = \sigma_p^{\rm SI}$. Hence, if the contribution from one of the $h_i$ lies on the diagonal lines and the contributions from the remaining mass eigenstates lie within the triangle, the SIDD cross section is dominantly mediated by that mass eigenstate. On the contrary, if the contributions lie outside the dashed diagonal lines, they interfere destructively to yield the total SIDD cross section. The left (right) panel shows parameter points where the lightest neutralino is bino (singlino) like. For both cases, we show points from our parameter scan which satisfy $\Omega h^2 = 0.12 \pm 50\,\%$.}
		\label{fig:DD_interference}
	\end{center}
\end{figure}

In Fig.~\ref{fig:DD_interference} we show the contributions to the SIDD cross section when taking into account only one Higgs mass eigenstate at a time as obtained from Eq.~\eqref{eq:SIDD} ignoring the sum over the CP-even Higgs mass eigenstates. We show these contributions plotted against the full SIDD cross section in the left (right) panel for mostly bino (singlino) points from our dataset satisfying all Higgs/collider constraints described above and featuring an acceptable relic density. For mostly bino DM, we find that SIDD cross sections as small as $\sigma^{\rm SI}_p \sim 10^{-13}\,$pb can be obtained by suppression of the coupling to the SM-like Higgs mass eigenstate alone. Destructive interference between different Higgs mass eigenstates is needed only for even smaller cross sections.

For mostly singlino DM, as shown in the right panel of Fig.~\ref{fig:DD_interference}, destructive interference between different Higgs mass eigenstates is almost always required to satisfy the experimental bounds on the SIDD cross section. This can be understood from the typical strength of singlino couplings $\sigma^{\rm SI}_p \propto g_{\widetilde{S}\widetilde{S} h}^2 \propto \lambda^4 \sim 0.1$, while binos couple with characteristic strength $\sigma^{\rm SI}_p \propto g_{\widetilde{B}\widetilde{B} h}^2 \propto g_1^4/4 \sim 0.004$. In addition, compared to bino DM, singlino DM has a much larger coupling to the scalar singlet state $H^S$ due to the presence of the tree-level coupling $\kappa$, cf. Eqs.~\eqref{eq:gSSS} and \eqref{eq:gBBS}. Hence we see the necessity of destructive interference between the contributions from the singlet like mass eigenstate $h_S$ and the SM-like mass eigenstate $h$ to suppress the SIDD cross section below the experimental limits.

Although blindspot cancellation or destructive interference arguably require some fine tuning, we stress that we readily find points in our dataset with SIDD cross sections below $\sigma_p^{\rm SI} \lesssim 10^{-13}\,$pb, out of reach of direct detection experiments for the foreseeable future. Such small cross sections are challenging to probe with current direct detection strategies due to the presence of the {\it neutrino floor}.

\begin{figure}
	\begin{center}
		\includegraphics[width=.49\linewidth]{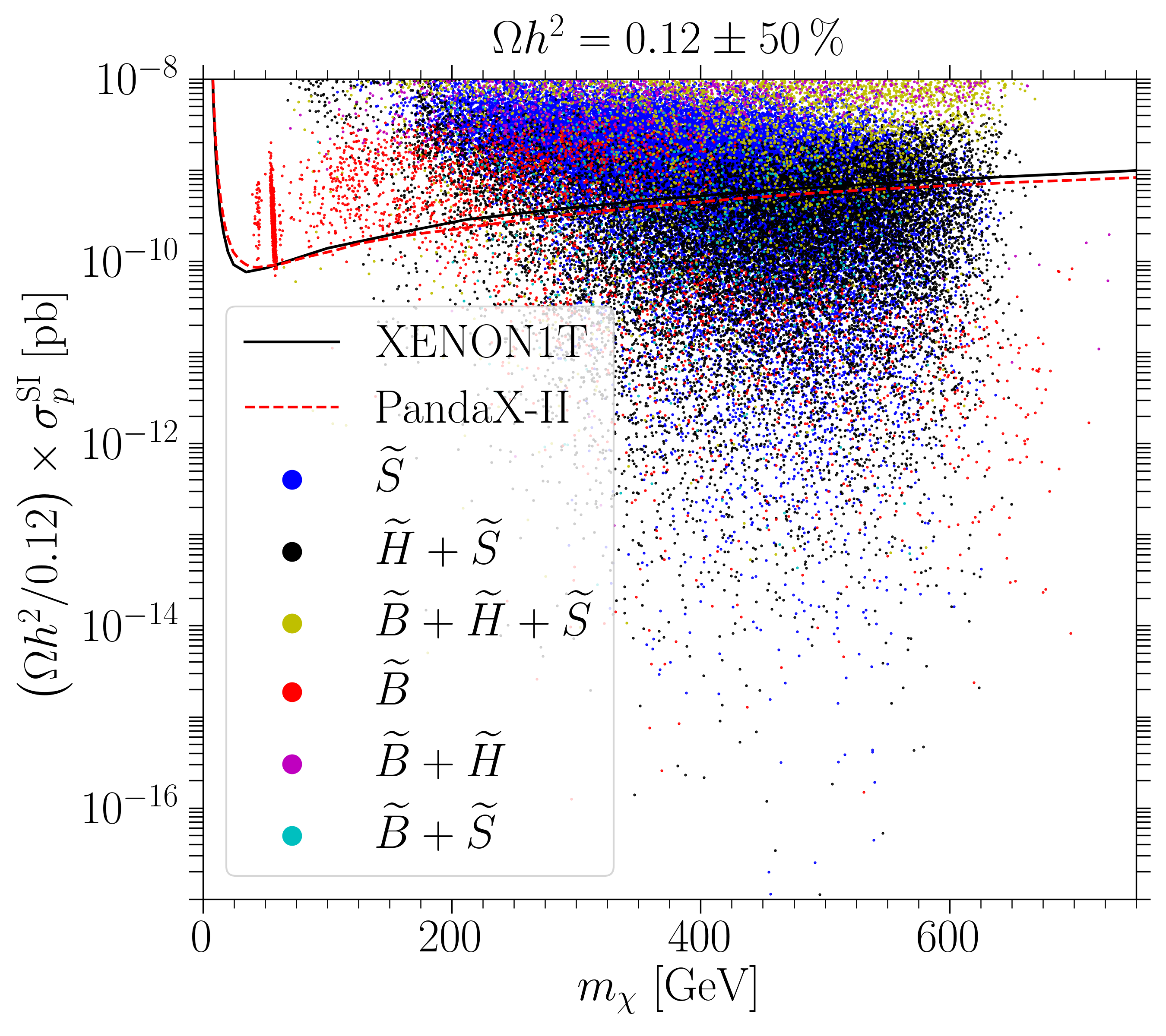}
		\includegraphics[width=.49\linewidth]{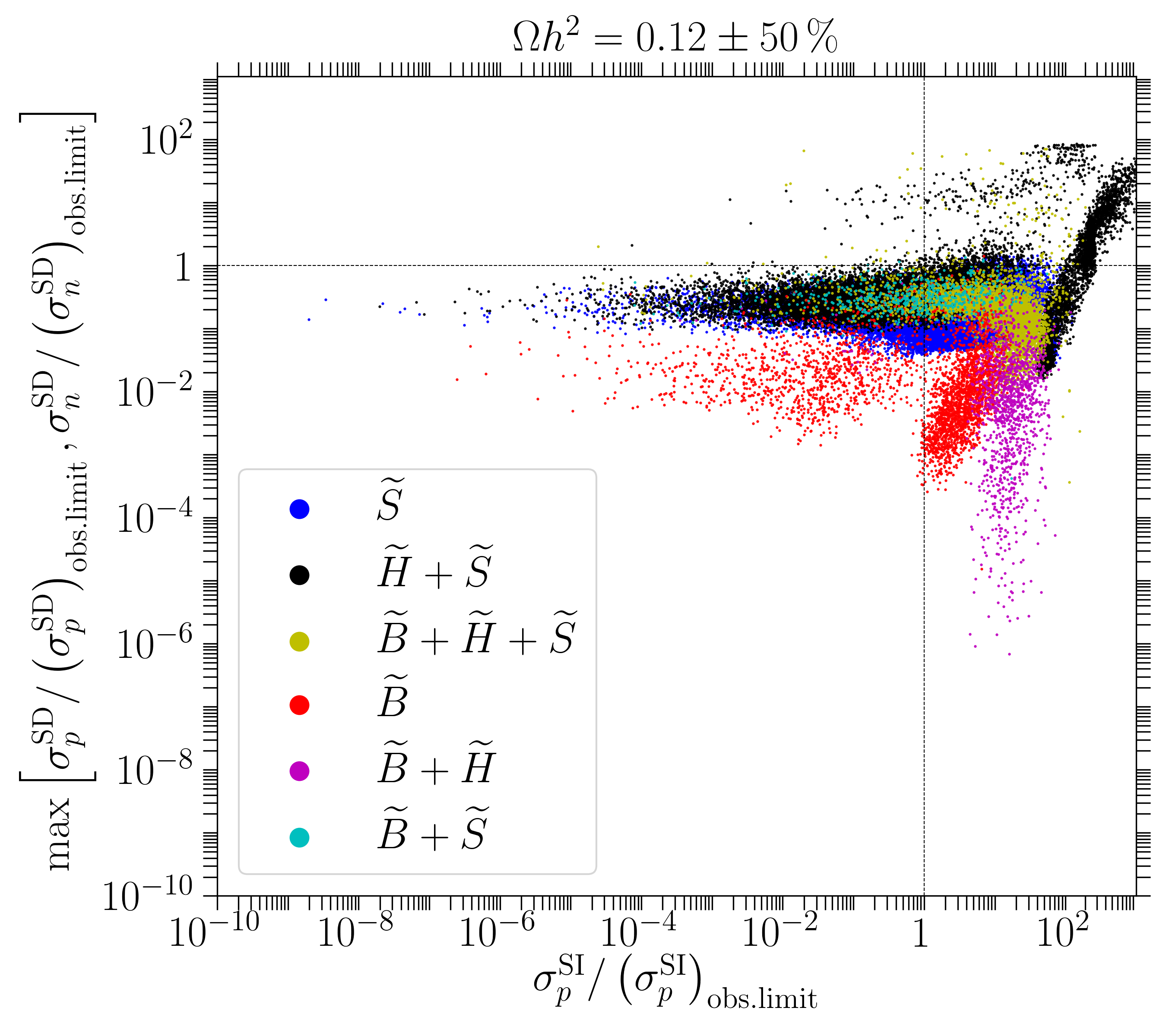}
		\caption{\textit{Left:} The SIDD cross section vs. the mass of the DM candidate $m_\chi$ for points from our parameter scan passing the required experimental collider constraints with acceptable relic density $\Omega h^2 = 0.12 \pm 50\,\%$. The solid black and dashed red lines indicate the most constraining experimental upper limits on the SIDD cross section from XENON1T~\cite{Aprile:2017iyp} and PandaX-II~\cite{Cui:2017nnn}, respectively. \textit{Right:} SIDD vs. SDDD cross section in units of the respective observed limit for the same points. For SIDD cross section, at each respective DM mass we use the stronger of the two limits from XENON1T and PandaX-II. For SDDD scattering we use the more constraining of the current bounds for either SDDD scattering of neutrons from LUX~\cite{Akerib:2017kat}, or SDDD scattering of protons from PICO-60~\cite{Amole:2017dex}. To guide the eye we indicate the current bounds with thin dashed lines; points lying in the lower left quadrant satisfy all current direct detection bounds. The color coding in both panels is the same as in Fig.~\ref{fig:DD_sigv_singlinos}.}
		\label{fig:DD_SI_SD}
	\end{center}
\end{figure}

In Fig.~\ref{fig:DD_SI_SD} we show the constraints from direct detection experiments for points from our scan with an acceptable relic density and satisfying collider constraints. The left panel shows the SIDD cross section vs. the DM mass. Note, that almost all parameter points with DM masses below the top mass $m_\chi < m_t \approx 173\,$GeV are ruled out by current SIDD constraints. The right panel shows a comparison of the constraining power of current SIDD and SDDD experiments~\cite{Aprile:2017iyp,Cui:2017nnn,Akerib:2017kat,Amole:2017dex} for the same points. As noted above, current experimental SIDD limits are much more constraining for our model than SDDD limits: as we can see in Fig.~\ref{fig:DD_SI_SD}, parameter points satisfying SIDD limits almost always satisfy current constraints on the SDDD scattering cross section from direct detection experiments, while the reverse is not true. Fig.~\ref{fig:DD_SI_SD} also shows that improvements in SDDD experiments will probe the remaining parameter space of our model more efficiently than a correspondingly large improvement in the bounds from SIDD. In particular, improving the sensitivity of SDDD scattering by approximately two orders of magnitude with respect to current bounds will probe most of the singlino region. A na{\"i}ve estimate of the sensitivity can be obtained by rescaling current bounds from LUX on the spin-dependent WIMP-neutron scattering cross section which were obtained with an exposure of $\epsilon = 129.5\,$kg\,yr~\cite{Akerib:2017kat}. The XENON1T experiment will have a sensitivity roughly one order of magnitude better assuming an exposure of $\epsilon = 1\,$t\,yr, while XENONnT could probe most of the singlino region assuming an exposure of $\epsilon = 20\,$t\,yr.

Finally, let us mention that out of the points passing the main phenomenological constraints described at the beginning of this section, which in particular lead to a SM-like Higgs boson compatible with LHC measurements and evade collider constraints as implemented in \texttt{NMSSMTools}, $\sim 15\,\%$ have an acceptable relic density $\Omega h^2 = 0.12 \pm 50\,\%$, and of the points with the correct relic density $\sim 35\,\%$ satisfy SDDD and SIDD constraints. 

\subsection{Indirect Detection}
In Section~\ref{sec:DMpheno} we have focused on direct detection constraints on our model. Any DM candidate is also subject to constraints from indirect detection experiments. Due to large astrophysical uncertainties on indirect detection constraints arising from charged particle production, we focus on constraints from photon emission here. The currently strongest indirect detection constraints on WIMP DM come from Fermi-LAT and MAGIC analyses of Milky Way satellite galaxies, ruling out canonical values of the thermally averaged annihilation cross sections $\langle \sigma v \rangle \sim 2 \times 10^{-26}\,{\rm cm}^3\,{\rm s}^{-1}$ for DM masses $m_\chi \lesssim 100\,$GeV if the annihilation is dominantly into pairs of $b$ quarks or $\tau$ leptons~\cite{Ahnen:2016qkx, Fermi-LAT:2016uux}. For DM masses $m_\chi \gtrsim m_t$ indirect detection bounds are much weaker, constraining thermal annihilation cross sections much larger than the typical values $\langle \sigma v \rangle_{x_F} \sim 2 \times 10^{-26}\,{\rm cm}^3\,{\rm s}^{-1}$ consistent with an acceptable relic density. Relevant constraints might further arise from the annihilation of DM matter particles captured via elastic scattering in the Sun, potentially giving rise to a flux of neutrinos observable at Earth. Under certain assumptions~\cite{Baum:2016oow}, most relevant for our case if DM annihilates dominantly into pairs of $W$ bosons, current bounds from DM capture and annihilation yield the most constraining upper limits on the spin dependent DM-proton scattering cross section~\cite{Tanaka:2011uf,Choi:2015ara,IceCube:2011aj,Aartsen:2016exj,Zornoza:2016xqb}. 

The parameter region in our model yielding an acceptable relic density and compatible with current SIDD constraints typically features DM masses heavier than the top quark $m_\chi \gtrsim m_t \approx 173\,$GeV. Furthermore, SIDD constraints require the coupling of DM to the SM-like Higgs boson to be suppressed which in turn implies that $(\chi\chi \to W^+ W^-)$ annihilation is suppressed. This is because of all the Higgs basis states, only $H^{\rm SM}$ couples to pairs of $W$ bosons. Thus, current indirect detection limits do not pose relevant constraints on our model since the annihilation modes most constraining for indirect detection bounds $(\chi\chi \to \bar{b}b / \tau^+ \tau^- / W^+ W^-)$ are suppressed for our preferred region of parameter space. Note however, that DM with mass $m_\chi > m_t$ decaying dominantly into pairs of top quarks may explain the excess of photons emitted from the Galactic Center observed by Fermi-LAT~\cite{Vitale:2009hr,Goodenough:2009gk,TheFermi-LAT:2017vmf,Freese:2015ysa}.

\subsection{Collider Constraints}
Our EFT DM model extends the neutral particle content of the SM by four Higgs mass eigenstates, two of them CP-even and two CP-odd, and a SM singlet Majorana fermion. The Dirac fermion we integrated out may be accessible at the LHC if its mass $\mu$ is below the TeV scale~\cite{Ellwanger:2016sur}. Note, that in our NMSSM ultraviolet completion we set the masses of all additional particles such as the SUSY partners of the SM fermions and the charged superpartners of the SM gauge bosons to values $\geq 2\,$TeV such that they are not directly accessible at the LHC.\footnote{Note, that this choice is made to allow for a straightforward connection of the NMSSM to our EFT model and allowing for lighter masses will generally not effect our analysis. Current experimental bounds from the LHC allow for much smaller masses of the SUSY particles, probeable at ongoing and future LHC runs.} Furthermore, DM is coupled to the SM via the Dirac fermion (the Higgsinos in the case of the NMSSM) which in turn couples to the Higgs and the $Z$ bosons. Thus, this model can be tested at the LHC either by DM pair production via $Z$ or Higgs bosons, or, via production of the additional Higgs bosons. 

As discussed previously in Sections~\ref{sec:EFT_Higgs} and~\ref{sec:NMSSM}, the observation of a 125\,GeV Higgs boson with couplings compatible with those of a SM Higgs~\cite{Aad:2015zhl,Khachatryan:2016vau} requires the presence of an $m_h \sim 125\,$GeV Higgs mass eigenstate approximately aligned with the $H^{\rm SM}$ Higgs basis interaction eigenstate. Direct detection limits furthermore strongly constrain the coupling of DM to both the SM-like Higgs and the $Z$ boson, such that the additional decay width of the $Z$ boson or the SM-like Higgs into DM particles must be small, even before taking into account the fact that $m_\chi \gg m_h/2 > m_Z/2$ in our model. The non-observation of signals from additional Higgs bosons at the LHC furthermore implies that additional Higgs bosons must either have masses heavier than $2 m_t \sim 350\,$GeV, or, be dominantly composed of the additional singlet interaction states $H^S$ or $A^S$ such that their production cross section is suppressed. A comparison of the constraining power of conventional searches for additional Higgs bosons decaying into pairs of SM particles can for example be found in the Appendix of Ref.~\cite{Baum:2017gbj}. Note, that most of the conventional Higgs searches such as $(gg \to H/A \to \tau^+\tau^- / \gamma\gamma / Z h)$ will only have small increases in sensitivity at future LHC runs because they are increasingly dominated by systematic errors. On the other hand, models with extended Higgs sectors and potentially large couplings between different Higgs mass eigenstates such as ours can be probed effectively by decays of heavy Higgs bosons into lighter Higgs bosons or a light Higgs and a $Z$ boson. Due to the presence of the SM-like Higgs, decays into pairs of SM-like Higgs bosons $(H \to h h)$ or into a SM-like Higgs boson and a $Z$ $(A \to Z h)$ are suppressed since the corresponding couplings are proportional to the mixing of the $H^{\rm SM}$ Higgs basis state with the $H^{\rm NSM}$ and $H^S$ states. If kinematically allowed, the branching ratios of $(H \to h h_S / Z a_S)$ and $(A \to Z h_S / h a_S)$ can however be sizeable. The collider signatures arising through such decays can be effective probes of our model~\cite{Carena:2015moc,Baum:2017gbj,Ellwanger:2017skc}.

\section{Conclusions} \label{sec:Conclusions}
Current bounds from direct detection experiments~\cite{Angloher:2015ewa,Agnese:2017jvy,Aprile:2017iyp,Cui:2017nnn,Akerib:2017kat,Amole:2017dex} have set stringent limits on WIMP DM scenarios where the relic density proceeds from thermal production mediated by Higgs and gauge bosons. In particular, direct detection experiments strongly constrain the coupling of the SM-like Higgs boson to DM particles and the vector-like coupling of DM to $Z$ bosons~\cite{Escudero:2016gzx}. The latter constraint is naturally satisfied for Majorana DM, since Majorana fermions couple to $Z$ bosons only via axial couplings. In this article, we explore an EFT describing the interactions of Majorana fermion WIMPs with an extended Higgs sector, comprised of a type II 2HDM and a SM gauge singlet. This model can be interpreted as an extension of Higgs portal models. In particular, we study the case where the EFT mass scale is identified with the mass of a heavy $SU(2)$-doublet Dirac fermion integrated out from the theory. Furthermore, we assume that all explicit mass terms are forbidden, which may be realized by imposing a $Z_3$ symmetry. In such a case the new fermions acquire mass through the vev of the singlet field. If this vev is generated by the same mechanism that induces electroweak symmetry breaking, we naturally obtain a weak scale mass for the DM candidate. In addition, the $Z_3$ symmetry reduces the number of allowed operators involving the singlet Majorana fermion and the doublet and singlet scalar fields, such that we can easily extend the EFT to dimension 6, including derivative operators. We derive the low energy Higgs spectrum and the couplings of DM to the neutral Higgs and gauge bosons in this model, and use these results to compute the relic density and spin independent direct detection cross section. We find blind spot conditions that allow for the evasion of spin independent direct DM detection constraints, while yielding the observed relic density. These blind spot conditions can be easily characterized in terms of the EFT parameters.

In order to test the validity of the results derived in our EFT model, we compare them to those obtained in the supersymmetric NMSSM, which features an analogous Higgs sector. We demonstrate that in the case of heavy gauge and fermion superpartners the NMSSM can be reduced to our EFT model at low energies, such that we can use it as an explicit computational basis. The Majorana singlet in our EFT is identified with the singlino in the NMSSM, and the EFT scale with the mass of the heavier Higgsinos. We show that the qualitative features obtained in the EFT are preserved in the complete theory, while the precise values of the couplings, relic density, and cross sections are modified by the presence of small corrections which could be included by dimension $d > 6$ operators in the EFT description. We also discuss the case of a $Z_3$ invariant Majorana fermion DM candidate, e.g. the bino in the NMSSM, and show that our EFT model can be generalized in a straightforward manner to include this scenario. 

Both in the EFT and in the NMSSM we show that the coupling of (singlino-like) DM to the SM-like Higgs is constrained to values below $g_{\chi\chi h} \lesssim 0.1$ by direct detection experiments, while simultaneously consistency with the DM relic density can be obtained through thermal annihilation via couplings of DM to the remaining Higgs and gauge bosons. The neutral Goldstone mode, comprising the longitudinal mode of the $Z$ boson, plays a prominent role in obtaining the thermal annihilation cross section. Moreover, in order to evade direct detection bounds, not only the coupling to the SM-like Higgs boson must be reduced, but in addition destructive interference of the SM-like Higgs boson with the singlet and/or non SM-like CP-even doublet Higgs boson is required to further suppress the cross section. 

In the NMSSM, when considering the case of light binos, we find a new {\it well tempered} DM region. For bino-like DM the couplings to the Higgs bosons tend to be smaller than for singlino-like one and direct detection bounds are hence evaded by a mild proximity to the blind spot conditions for the DM coupling to the SM-like Higgs state. The relic density in the {\it well tempered region} is obtained via co-annihilation of the bino with the singlino. Beyond the {\it well tempered region}, the correct relic density for bino-like DM may also be obtained via resonant annihilation with either the $Z$ boson or a Higgs mass eigenstate.

In both the cases of bino-like and singlino-like DM, we find that the current constraints coming from SDDD are subdominant compared to those
coming from SIDD. However, we also find that an improvement of two orders of magnitude in the bounds on the SDDD cross sections could efficiently
probe most models consistent with a blind spot in SIDD. 

Finally, collider searches for the Majorana DM particles are mostly restricted to the usual channels of single SM-particles plus missing energy, in particular those associated with Higgs bosons. Going beyond the effective theory, these searches may be complemented by missing energy searches in the production and decay of the Dirac doublet fermion and, in the NMSSM, by related searches for heavy superpartners.

\acknowledgments

We would like to thank Bibhushan Shakya for interesting discussions related to this work. SB acknowledges support from the Swedish Research Council (Vetenskapsr\r{a}det) through the Oskar Klein Centre (Contract No. 638-2013-8993). NRS is supported by Wayne State University. This manuscript has been authored by Fermi Research Alliance, LLC under Contract No. DE-AC02-07CH11359 with the U.S. Department of Energy, Office of Science, Office of High Energy Physics. The United States Government retains and the publisher, by accepting the article for publication, acknowledges that the United States Government retains a non-exclusive, paid-up, irrevocable, world-wide license to publish or reproduce the published form of this manuscript, or allow others to do so, for United States Government purposes. Work at University of Chicago is supported in part by U.S. Department of Energy grant number DE-FG02-13ER41958. Work at ANL is supported in part by the U.S. Department of Energy under Contract No. DE-AC02-06CH11357. The work of MC, NS and CW was partially performed at the Aspen Center for Physics, which is supported by National Science Foundation grant PHY-1607611.

\newpage

\appendix
\section{EFT Lagrangian} \label{app:L}

Rescaling the Majorana fermion field $\chi$ in order to retain a canonical kinetic term and keeping terms to order $\mathcal{O}(\mu^{-2})$, the EFT Lagrangian~\eqref{eq:EFTmu} reads in terms of the Higgs basis states, Eqs.~\eqref{eq:Hbasis1}-\eqref{eq:Hbasis-1},
\begin{equation} \begin{split} \label{eq:EFTmuHbasis}
	\mathcal{L} =&~ \frac{\delta}{2 \mu} \chi\chi \left\{ \frac{s_{2\beta}}{2} \left[ 2 v^2 + 2\sqrt{2} v H^{\rm SM} + (H^{\rm SM})^2 - (H^{\rm NSM})^2 - (A^{\rm NSM})^2 + (G^0)^2 \right] \right.
	\\ & \left. \qquad + c_{2\beta} \left[ \sqrt{2} v H^{\rm NSM} + H^{\rm SM} H^{\rm NSM}+ A^{\rm NSM} G^0 \right] \right.
	\\ & \left. \qquad + i \left[ \sqrt{2} v A^{\rm NSM} + H^{\rm SM} A^{\rm NSM} - H^{\rm NSM} G^0 \right] \right\} + {\rm h.c.}
	\\ & \quad - \frac{\lambda \delta}{2 \sqrt{2} \mu^2} \chi\chi \left( H^S + i A^S \right)
	\\ & \qquad \times \left\{ \frac{s_{2\beta}}{2} \left[ 2 v^2 + 2\sqrt{2} v H^{\rm SM} + (H^{\rm SM})^2 - (H^{\rm NSM})^2 - (A^{\rm NSM})^2 + (G^0)^2 \right] \right.
	\\ & \qquad \left. + c_{2\beta} \left[ \sqrt{2} v H^{\rm NSM}+ H^{\rm SM} H^{\rm NSM} + A^{\rm NSM} G^0 \right] \right.
	\\ & \qquad \left. + i \left[ \sqrt{2} v A^{\rm NSM}+ H^{\rm SM} A^{\rm NSM}- H^{\rm NSM} G^0 \right] \right\} + {\rm h.c.}
	\\ & \quad - \left(\kappa - \frac{\kappa \alpha v^2}{|\mu|^2} \right) \left[ \frac{\mu}{\lambda} + \frac{1}{\sqrt{2}} \left( H^S + i A^S \right) \right] \chi\chi + {\rm h.c.}
	\\ & \quad - \frac{\kappa \xi}{2|\mu|^2} \left[ \frac{\mu}{\lambda} + \frac{1}{\sqrt{2}} \left( H^S + i A^S \right) \right] \chi\chi
	\\ & \qquad \times \left[ 2 v^2 + 2\sqrt{2} v H^{\rm SM} + (H^{\rm SM})^2 + (G^0)^2 + (H^{\rm NSM})^2 + (A^{\rm NSM})^2 \right] + {\rm h.c.}
	\\ & \quad + \frac{\alpha}{|\mu|^2} \frac{g_1}{2 s_W} Z_\mu \chi^\dagger \bar{\sigma}^\mu \chi 
	\\ & \qquad \times \left\{ \frac{c_{2\beta}}{2} \left[ - 2 v^2 - 2\sqrt{2} v H^{\rm SM} - (H^{\rm SM})^2 + (H^{\rm NSM})^2 + (A^{\rm NSM})^2 - (G^0)^2 \right] \right.
	\\ & \qquad \left. + s_{2\beta} \left[ \sqrt{2} v H^{\rm NSM} + H^{\rm SM} H^{\rm NSM} + A^{\rm NSM} G^0 \right] \right\} 
	\\ & \quad + \frac{\alpha}{|\mu|^2} \chi^\dagger \bar{\sigma}^\mu \chi \left[ i s_{2\beta} \left( \frac{v}{\sqrt{2}} i \partial_\mu A^{\rm NSM} + H^{\rm SM} i \partial_\mu A^{\rm NSM} + H^{\rm NSM} i \partial_\mu G^0 \right) \right.
	\\ & \qquad \left. - i c_{2 \beta} \left( \frac{v}{\sqrt{2}} i \partial_\mu G^0 + H^{\rm SM} i \partial_\mu G^0 - H^{\rm NSM} i \partial_\mu A^{\rm NSM} \right) \right] 
	\\ & \quad + \frac{\alpha}{2 |\mu|^2} \chi^\dagger i \bar{\sigma}^\mu \partial_\mu \chi \left[ 2\sqrt{2} v H^{\rm SM} + (H^{\rm SM})^2 + (G^0)^2 + (H^{\rm NSM})^2 + (A^{\rm NSM})^2 \right].
\end{split} \end{equation}
The couplings of the Majorana fermions to the Higgs basis states and the $Z$ boson can be read off from here. Note, that couplings $g_{\chi \chi \Phi_i \Phi_j \ldots}$ to Higgs basis states written explicitely in this work are normalized as
\begin{equation}
	g_{\chi \chi \Phi_i \Phi_j \ldots} \equiv \frac{\partial \mathcal{L}}{\partial \chi^2 \, \partial \Phi_i \, \partial \Phi_j \, \dots} \;,
\end{equation}
while couplings involving $Z$ bosons are normalized as
\begin{equation}
	g_{\chi \chi Z \Phi_i \ldots} \equiv \frac{1}{2} \frac{\partial \mathcal{L}}{\partial \chi^2 \, \partial Z \, \partial \Phi_i \, \dots} \;.
\end{equation}

\bibliographystyle{JHEP.bst}
\bibliography{nmssmbib}

\end{document}